\documentclass{aa}
\usepackage[varg]{txfonts}
\usepackage[utf8]{inputenc}

\usepackage{hyperref}
\usepackage{graphicx}
\usepackage{amsmath}
\usepackage{amssymb}
\usepackage{txfonts}
\usepackage{color}
\usepackage{multirow}

\usepackage{aas_macros}
\usepackage{natbib}
\bibpunct{(}{)}{;}{a}{}{,}
\hypersetup{colorlinks,citecolor=blue}

\begin{document}
    
\title{Thermal instability as a constraint for  warm X-ray corona in AGN}

\author{Dominik Gronkiewicz \inst{1}
\and Agata Różańska\thanks{E-mail: agata@camk.edu.pl (AR)} \inst{1}
\and 
Pierre-Olivier Petrucci \inst{2}
        \and
        Renaud  Belmont \inst{3}
         }
         
 \offprints{A. Różańska}         

\institute{Nicolaus Copernicus Astronomical Center, Polish Academy of Sciences, Bartycka 18, 00-716 Warsaw, Poland  
\and 
	Universit{\'e} de Grenoble Alpes, IPAG, F-38000 Grenoble, France 
	\and
       Université Paris Cité, Université Paris-Saclay, CEA, CNRS, AIM, F-91191, Gif-sur-Yvette, France
}

\date{Received 2 November 1992 / Accepted 7 January 1993}


\abstract
    {Warm corona is a possible explanation for Soft X-ray Excess in Active Galactic Nuclei (AGN). This paper contains self consistent modeling of both: accretion disk with optically thick corona, where the gas is heated by magneto-rotational instability dynamo (MRI), and cooled by radiation which undergoes free-free absorption and Compton scattering.} 
{We determine the parameters of warm corona in AGN using disk-corona structure model that takes into account magnetic and radiation pressure. We aim to show the role of thermal instability  (TI) as a constraint for warm, optically thick X-ray corona in AGN.} 
{With the use of relaxation code, the vertical solution of the disk driven by MRI together with radiative transfer in hydrostatic and radiative equilibrium is calculated, which allows us to point out how TI affects the corona for wide range of global parameters.} 
{We show that magnetic heating is strong enough to heat upper layers of the accretion disk atmosphere, which form the warm corona covering the disk. Magnetic pressure does not remove TI caused by radiative processes  operating in X-ray emitting plasma. TI disappears only in case of accretion rates  higher than 0.2 of Eddington, and high magnetic field parameter $\alpha_{\rm B} > 0.1$.}
{TI plays the major role in the  formation of the warm corona above magnetically driven accretion disk in AGN. 
 The warm, Compton cooled corona, responsible for soft X-ray excess, resulted from our model has typical temperature 
 in the range of 0.01 - 2 keV and optical depth even up to 50, which agrees with recent observations.}


\maketitle


\section{Introduction}

Soft X-ray excess is commonly observed in large majority of active galactic nuclei (AGN) 
\citep[e.g.][]{1987-Pounds-Mkn335,walter93,1998-Magdziarz,
page2004,gierlinski2004,bianchi2009,2011-Mehdipour-Mrk509,2012-Done-SXE,2013-Petrucci-Mrk509,2016-Keek,2018-Petrucci} including 
quasars \citep{1988-Madau,1994-Laor-1,1997-Laor-2,2004-GierlinskiDone,2005-Piconcelli-SXE}.
It appears as an excess in emission,
when extrapolating the 2-10\,keV power law of AGN { to the soft X-ray band}. 

The origin of this spectral component is still under debate, but two major scenarios are currently considered. The first scenario holds that soft X-ray excess 
{
is a result of blurred ionized reflection \citep{crummy2006,walton2013,2019ApJ...871...88G}, 
but such model produces many lines that are not observed in the soft X-ray band,
and extreme blurring is generally required to wash them out. 
Recently, detailed radiative transfer analysis in the warm corona, proved that those lines cannot be easily created due to the  full domination of  internal heating and Compton scattering  over the line transitions \citep{petrucci2020,ballantyne2020}.
}

Second scenario relies on the fact, that soft X-ray excess is produced due to the Comptonization in optically thick  ($\tau \sim 10-20$) warm corona, which is reasonable assumption while fitting the data \citep[e.g.][and references therein]{1998-Magdziarz,jin12,2013-Petrucci-Mrk509,2018-Porquet,2018-Petrucci}.
Nevertheless, from data fitting, both above scenarios require the existence of the warm layer which is additionally heated, not only by radiation, but also by mechanical process, and the warm layer is located next to the accretion disk. 
 On the other hand, none of those models consider
 how this warm layer is physically caused, and only the constant 
heating over gas volume of the warm corona is assumed and included in model computations as a free parameter \citep{petrucci2020,ballantyne2020}.

Analytically, we have predicted that high optical depth and high 
temperature of the warm, Compton cooled corona is possible when additional pressure component and mechanical heating are taken into account \citep{rozanska15}.
Recently, we extended this analytical model by numerical calculations of magnetically heated disk with free-free processes taken into account in addition to Compton scattering in case of the accretion disk around black hole of stellar mass i.e. Galactic black hole binaries (GBHB)
\citep[][hereafter GR20]{2020-Gronkiewicz}.
For such sources, an accretion disk is already quite hot, visible in X-rays, and slightly warmer  layer can form on different distances from black hole. In GR20, we shown
that the warm corona, which is physically fully coupled with the cold,
{ magnetically supported disk} (MSD), arises naturally due to magnetic heating.
Such corona is cooled mostly by Compton scattering, in agreement with the observations.
We clearly demonstrated that the classical thermal instability (TI) \citep{field1965,1996-RozanskaCzerny}, caused by gas cooling processes, cannot be removed by magnetic pressure, and it shapes the radial distance of the warm corona in those objects. 

But the case of AGN is different, since the disk temperature is relatively cool, of the order of $10^{4-5}$\,K, and the transition to the warm, $10^{7}$\,K, X-ray corona is accompanied by strong changing of gas global parameters, as density and pressure. On the other side, the physically consistent model of warm corona responsible for soft X-ray excess is desirable as we have the growing collection of observed sources indicating such feature, obtained  with modern X-ray satellites. 
In this paper, we adopt GR20 model to the case of AGN, where the  geometrically thin and optically thick accretion disk is around the supermassive  black hole (SMBH). 
We assume  that the accretion disk is magnetized and the magneto-rotational instability (MRI) is the primary source of viscosity and energy dissipation.
We directly use the analytic formula derived  by \cite[][hereafter BAR15]{2015-Begelman}, where the vertical profile of magnetic heating of the accretion flow is determined.
On the top of this assumption, the disk vertical structure together with the radiative transfer equation in gray atmosphere are fully solved with the relaxation method proposed originally by \citep{Henyey1964}. 

As a results of our computations, we obtain the optical depth and temperature of the warm corona in AGN, which are the main observable 
when analyzing X-ray data. { The specific heating of a warm corona formed in our model, 
is computed self consistently based on MRI and reconnection. The value of this heating together with the magnetic pressure, change in the vertical direction, which is a big improvement of the 
constant heating warm corona slab used in recent calculations of energy spectrum from warm corona \citep{petrucci2020,ballantyne2020}. The next improvement of our analytical model \citep{rozanska15} is the use of free-free radiative process which is crucial for TI to appear. 
We show, that in case of AGN, the magnetic pressure is high enough to produce stable 
brunch in classical TI curve obtained under constant gas and radiation pressure condition.  
MSD with warm corona is dominated by magnetic pressure 
which makes radiative cooling gradient always positive when computed under constant gas plus magnetic pressure. Therefore, initially thermally unstable zone, through which the radiative cooling gradient under constant gas pressure is negative, is frozen in the magnetic field allowing corona to exist}. All results obtained by us, are 
compared with observations of unabsorbed type 1 AGN 
\citep[][and references therein]{jin12,2013-Petrucci-Mrk509,2018-Ursini,2019-Middei}, and with models computed in case of GBHB (GR20). We show, that 
the measured parameters agree with those obtained by our numerical computations. 
Our solutions are prone to TI in a wide range of global parameters, indicating that
 TI plays a crucial role in soft corona formation in AGN.

The structure of the paper is as follows: 
Sec.~\ref{sec:met} presents the method of our computations, while 
model parameters in case of AGN are given in Sec.~\ref{sec:params}.
The implementation of MRI-quenching and the role of magnetic pressure 
is separately discussed in Sec.~\ref{sec:mriq}.
The radiation pressure supported solution is analyzed in 
Sec.~\ref{sec:inflated}. 
For comparison with observations we used models generated in our scheme of random choice described in Sec.~\ref{sec:ran} and measurements described in 
Sec.~\ref{obs:pred}. Results of our numerical computations are 
presented in Sec.~\ref{sec:res}, where 
we display vertical structure of disk/corona system, but we also show radial limitation for which optically thick corona can exist.
Discussion and conclusions are given in Sec.~\ref{sec:dis} and \ref{sec:con}  respectively.


\section{Set-up of the model}
\label{sec:met}

We further investigate the model developed by GR20, { adapted} for the case of AGN, where an accretion disk is around SMBH.
The code solves the vertical structure of a stationary, optically thick accretion disk,
{
with radiation transfer assuming gray medium and temperature determined by solving balance between total heating and cooling including magnetic reconnection and radiative processes. The gas heating is powered by the magnetic field, according to the analytic formula given by BAR15, and the disk is magnetically supported over it's whole vertical extent. As an output, we can determine the vertical profile of magnetic heating of the accretion disk, as well as of the corona that forms on the top of the disk. 
}
{ For the purpose of this paper we consider Compton scattering and free-free emission/absorption as radiative processes which balance the magnetic heating.}
Our approach is innovative because it 
connects the magnetically supported disk (BAR15) with transfer of radiation in optically thick and warm medium, in thermal equilibrium \citep{rozanska15}.
Such approach allows us to verify if the warm corona can be formed above MSD, and stay there in optically thick regime, which will be consistent with many observational cases. 

The following set of five  equations is solved:  

\begin{equation}
 z \frac{d P_{\rm mag}}{dz} + \left( 2 + \frac{\alpha_{\rm B} \nu}{\eta} \right) P_{\rm mag} - \frac{\alpha_{\rm B}}{\eta}  \left( 
P + P_{\rm mag} \right) = 0,
\end{equation}

\begin{equation}
{\cal H} = 2 (\eta + \alpha_B \nu) \Omega P_{\rm mag} - \alpha_B \Omega \left( 
P + P_{\rm mag} \right) = \Lambda (\rho, T, T_{\rm rad}),
\end{equation}

\begin{equation}\label{eq:dfraddz}
\frac{d F_{\rm rad}}{dz} = \Lambda (\rho, T, T_{\rm rad}),
\end{equation}

\begin{equation}\label{eq:dtraddz}
\frac{dT_{\rm rad}}{dz} + \frac{3 \kappa \rho}{16 \sigma T_{\rm rad}^3} F_{\rm rad} = 0,
\end{equation}

\begin{equation}\label{eq:hydrostatic}
\frac{d P_{\rm gas}}{dz} + \frac{d P_{\rm mag}}{dz} + \rho  \left[ \Omega^2 z - \frac{\kappa F_{\rm rad}}{c} \right] = 0,
\end{equation}
where the first two equations describe the magnetic field structure and 
magnetic heating as presented by BAR15. Magnetic pressure vertical gradient
$dP_{\rm mag}/dz$ depends on magnetic parameters: 
total magnetic viscosity - $\alpha_{\rm B}$, magnetic buoyancy parameter - $\eta$, and reconnection efficiency parameter - $\nu$ (see: GR20, for definition), and the sum of gas and radiation pressure 
$P= P_{\rm gas} + P_{\rm rad}$.  { At each point of disk/corona vertical structure}, 
the local magnetic heating ${\cal H}$ is balanced by net radiative cooling rate 
$\Lambda (\rho, T, T_{\rm rad})$, { where the latter is calculated from the frequency-integrated radiative transfer equation (see GR20 for exact formulae)}. 

Equations \eqref{eq:dfraddz} and \eqref{eq:dtraddz}
describe how radiation flux is locally generated and transported by the gas. 
The fifth equation is the momentum equation in stationary situation i.e. hydrostatic equilibrium, where $\rho$ is gas density, $\kappa$ is Rosseland mean opacity,
$\Omega$ is Keplerian angular velocity, 
and $c$ is the light velocity. The gas and radiation pressure in our model can locally be described 
as $P_{\rm gas} = \frac{k}{\mu m_{\rm H}} \rho T$ and 
$P_{\rm rad} = \frac{4 \sigma}{3c} T_{\rm rad}^4$ which is typical in case of 
gray atmosphere with $k$ being Boltzmann constant, $\sigma$ - Stefan-Boltzmann constant, and  $\mu$ - mean molecular weight.

To formulate our net radiative cooling function, 
we assume Compton electron scattering and free-free absorption/emission as radiative processes that occur in the medium. 
The radiative cooling function { is calculated at each point of the MSD's vertical structure} and in this paper has the form:
\begin{equation}\label{eq:lcoolambda}
\Lambda  \left( \rho, T, T_{\rm rad} \right) \equiv  4 \sigma \rho  \left[  \kappa_{\rm ff} \left( T^4 - T_{\rm rad}^4 \right)  + \kappa_{\rm es} T_{\rm rad}^4 \frac{ 4k\left(\gamma T - T_{\rm rad} \right) }{ m_{\rm e} c^2 } \right],
\end{equation}
where $\gamma = 1 + 4kT/(m_{\rm e} c^2)$ accounts for relativistic limit in Compton cooling, $\kappa_{\rm es}$ is electron scattering opacity and $\kappa_{\rm ff}$ is Planck-averaged free-free opacity.
In principle, more processes can be included \citep{1999-Rozanska},
as long as Planck-average opacities are known. However, in this paper we clearly show that free-free absorption is efficient enough to onset of TI on the particular optical depth in magnetically supported accretion disk atmosphere. Usually, TI is connected with ionization and recombination processes, which are not taken by us into account. We show here that ionization is not necessary to study TI, but it will be valuable to test in the future how ionization/recombination processes influence the width of an thermally unstable region, 
which as we show below is connected with strength of warm corona. We plan to do it in our future work, since it requires substantial extension of our numerical code. 

The set of equations and the associated boundary conditions are solved using a relaxation method where the differential equations are discretized and then iteratively solved for convergence \citep[][GR20]{Henyey1964}. { We adopt typical boundary conditions appropriate for a cylindrical geometry of an accretion disk i.e.: radiative flux at the mid-plane must be zero, but there is non-zero magnetic pressure at the equatorial plane. 
The value of magnetic parameter at the disk mid-plane $\beta_0=P_{\rm gas}/P_{\rm mag}$ 
{can be derived from  three magnetic input parameters, described} 
in Sec.~\ref{sec:params} below.
At the top of atmosphere of MSD, we assume that the sum of the flux 
carried away by radiation and of the magnetic field is equal to the flux obtained 
by Keplerian disk theory. We also take into account standard boundary condition for radiative transfer, i.e. mean intensity $J=2H$, where Eddington flux $H$ connects to the radiative flux as $4 \pi H=F_{\rm rad}$.}
For full numerical procedure and boundary conditions of our code we refer the reader 
to GR20 paper Appendix A. 
The numerical program in FORTRAN, Python and Sympy we developed, is available online\footnote{\texttt{http://github.com/gronki/diskvert}}.


\subsection{Model parameters}
\label{sec:params}

The model is parameterized by six parameters. The first three are associated with an accretion disk and they are: black hole mass $M_{\rm BH}$, accretion rate $\dot m$ and distance from the nucleus $R$. Wherever not specified, we assume $M_{\rm BH} = 10^8$\,M$_{\odot}$, $\dot m = 0.1$ in the units of Eddington accretion rate, and $R = 6 R_{\rm Schw}$, where $R_{\rm Schw}= 2GM/c^2$, with $G$ being the gravitational constant, as our canonical model named: CM. { Those parameters were chosen only for the representation of our results, nevertheless, we are able to compute the models for other typical parameters within the broad range.
} 

The next three parameters, which are: magnetic viscosity - $\alpha_{\rm B}$ , magnetic buoyancy parameter - $\eta$, and  reconnection efficiency parameter - $\nu$,   
define the local vertical structure of MSD. { We know how those parameters depend on each other, but there is no one good way of determining what values they should take.
In our previous paper GR20, we made an effort to compare magnetic structure with 
the one obtained from MHD simulations by \cite{2016-Salvesen}, and we adjusted those parameter relations to fit the simulation results. 
However, this is only one of the assumptions that can be made about the relation between $\alpha_{\rm B}$, $\eta$ and $\nu$, and in this paper we choose to follow a slightly different approach.}
First, we require that the vertically averaged ratio of 
magnetic torque over magnetic pressure, marked as $A$,  should be roughly constant in the disk, consistent with { simulations reported by} \cite{2014-Jiang} and \cite{2016-Salvesen}.
It can be proven that this is true if the value $A$ is constant for any $\alpha_{\rm B}$, since
\begin{equation}
    t_{r \phi} \Omega^{-1} = \alpha_{\rm B} P_{\rm tot} = \alpha_{\rm B} \frac{P_{\rm tot}}{P_{\rm mag}} P_{\rm mag}  = 2 A P_{\rm mag} ,
    \label{eq:mvis}
\end{equation}
where
\begin{equation}
    \label{eq:consta}
    A = \frac{1}{2} \alpha_{\rm B} \nu + \eta .
\end{equation}
With this information we can simplify our parametrization of the magnetic torque by 
making $\eta$ and $\nu$ depend on $\alpha_{\rm B}$, since the simulations by \cite{2016-Salvesen} show that there is some correlation between three parameters for different magnetic field strengths (see: Fig.2 in GR20).

We first eliminate $\eta$ by  introducing the constant $p$, and 
assuming that it depends on $0 < \alpha_{\rm B} < 2A$ as follows:
\begin{equation}
    \label{eq:eta}
    \eta \left( \alpha_{\rm B} \right) = A \left( \alpha_{\rm B} \over 2A \right)^p .
\end{equation}
Next, we transform the relation in Eq.~\ref{eq:consta} and substitute the expression in Eq.~\ref{eq:eta} to obtain $\nu$
\begin{equation}
    \nu \left( \alpha_{\rm B} \right) = 2\frac{A - \eta}{\alpha_{\rm B}} = \frac{1 - \left( \alpha_{\rm B} \over 2A \right)^p}{{\alpha_{\rm B} \over 2A }}.
    \label{eq:nu}
\end{equation}
{ The above relations were our assumption, and were formulated in the process of understanding what really influences the magnetic heating vertical structure. For instance, we have tested that our results are not sensitive to the selection of $A$ and $p$, therefore we keep those values constants for the results of this paper i.e. $A = 0.3$ and $p = 0.3$.}

{ Therefore, keeping constant $A$ and $p$, the two extremes of our magnetic viscosity parameter 
range are realized where $\alpha_{\rm B} \approx 2A$, which is a strongly magnetized disk, and where 
$\alpha_{\rm B} \approx 0$ for weakly magnetized disk (Eq.~\ref{eq:mvis}).
Furthermore, magnetic buoyancy parameter - $\eta$ and reconnection efficiency parameter - $\nu$, can be automatically provided by Eqs.~\ref{eq:eta} and 
\ref{eq:nu} for an assumed value of $\alpha_{\rm B}$. Such convention may seem a bit complicated, but 
it gives us an intuition about the values of the efficiency of other processes as magnetic buoyancy strength and reconnection efficiency. In case of strongly magnetized disk, we obtain 
$\eta \approx A $ and $\nu \approx 0$, which means that such disk has both: very little dynamo 
polarity reversals and low recconection efficiency. Weakly-magnetized disk yields to 
$\eta \approx 0 $ and $\nu \gg 1$ and it is realized when frequent polarity reversals cause
majority of the energy to be dissipated by magnetic reconnection.}


\subsection{Implementation of MRI-quenching}

\label{sec:mriq}

The vertical support in the disk is caused by the toroidal field produced by MRI dynamo in the presence of vertical net energy field, and the efficiency of this process is described by 
the magnetic viscosity parameter $\alpha_{\rm B}$.
However, when the magnetic field becomes too strong compared to kinetic force of the gas, the MRI is inefficient, and the dynamo is quenched (BAR15).
That causes the decrease in toroidal field production in some areas of the vertical structure and reduces the magnetic support.
The approximate condition for this efficiency limit is given by \cite{2005-PessahPsaltis} and can be written as a limit to magnetic pressure:
\begin{equation}\label{eq:pmagmax}
P_{\rm mag} \leq P_{\rm mag,max} = \sqrt{ \frac{5}{3} \rho P_{\rm gas}  }  \,\, \Omega R.
\end{equation}
Since the limit on the magnetic pressure depends on density and gas pressure, it becomes particularly significant in AGN disks, where the contribution of gas pressure compared to radiation pressure is lesser than in stellar-mass black hole accretion disks, i.e. the domination of radiation pressure happens for larger ranges of accretion rates and extends for larger distances from black holes 
\citep{2008-Kato}. 

We implement this condition by limiting the $\alpha_{\rm B}$ parameter:
\begin{equation}%
\label{key}
\alpha_{\rm B}' = \alpha_{\rm B} \,\, T(P_{\rm mag,max} / P_{\rm mag}) =  \alpha_{\rm B}  \,T(x),
\end{equation}
where $T(x) = x^4 / (1 + x^4)$ is a smooth threshold-like function with values close to $0$ for $x < 1$ and close to $1$ for $x \gg 1$.
Using this method, we are able to obtain a gradual transition between the zones where
MRI dynamo operates and zones where it is quenched.
We have checked that the final results of our model are not sensitive on the choice of 
the particular shape of threshold function.

{ The magnetic field was never measured from accretion disk around SMBH, and therefore, we have no constrains how it should be. Only simulations give us some clues and we have check that our resulted vertical magnetic field distribution agrees with simulations. (GR20). Even if BR15 model is assumed, the energy released in magnetically heated disk has the comparable value to the energy generated by viscosity, and presents a nice way of energy production by MRI and the transfer of this energy into gas. Below we show, how this magnetic pressure affects radiative processes in accretion disks around SMBH.} 

\subsection{Thermal instability}

To determine the temperature of the gas, we need to solve the thermal balance equation, where on one side we have heating terms, and on the other side the radiative cooling function. 
This equation normally has one solution, and the cooling rate should have a positive derivative with respect to temperature.
In some circumstances this problem has more than one solution, typically three, one of them being unstable.
For that unstable solution, assuming that the heating is roughly independent on gas parameters, the condition for instability reads
\begin{equation}
\frac{d \ln \Lambda}{d \ln T}
= \frac{\partial \ln \Lambda}{\partial \ln T}
- a \frac{\partial \ln \Lambda}{\partial \ln \rho}
 \leq 0,
\end{equation}
where $a$ is a numerical constant, depending on the constraining regime.
Some typical values are: $a = 1$ for $\delta P_{\rm gas}=0$, $a = 0$ for $\delta \rho = 0$ and $a = \frac{\beta}{\beta + 2}$ for $\delta P_{\rm gas} + \delta P_{\rm mag} = 0$ where $\beta = P_{\rm gas} / P_{\rm mag}$ \citep{field1965}.

Usually, $\delta P_{\rm gas}=0$ is the assumed thermodynamic constraint for the instability, and
if only Compton scattering and free-free cooling are taken into account, this limits the density of the gas, according to formula given in GR20:
\begin{equation}
n / n_0 < (T / T_{\rm rad})^{-1/2},
\end{equation}
where
\begin{equation}\label{eq:n0}
n_0 
= 4 \times 10^{16} \left( \frac{T_{\rm rad}}{10^6 {\rm K}} \right)^{9/2} \,.
\end{equation}
If we assume that around the base of the corona $T \approx T_{\rm rad}$, this condition becomes $n < n_0$, and we can estimate the approximate limit for optical depth of the corona.

The estimate for density given in Eq.~\ref{eq:n0} assumes that gas pressure is constant.
In magnetically-dominated corona, however, the density in hydrostatic equilibrium is mostly imposed by magnetic field gradient.
For that case, we might assume that the constant-density model is more relevant than the constant-pressure model.
Luckily, in isochoric regime the condition for instability is never satisfied, and when magnetic pressure is added, even small values are enough to completely eliminate the instability.  

\subsection{Radiation-pressure supported solutions}
\label{sec:inflated}

For SMBHs, as opposed to the black holes of stellar mass, the radiation pressure has a large contribution in comparison to the gas pressure in supporting the disk structure. 
This is important because radiation pressure dominated disks are considered unstable for  
the certain range of parameters  \citep{1974-Lightman,1975-Shibazaki,1976-Shakura,2006-Begelman-Bubbles,2011-Janiuk}.
When solving the vertical structure of a geometrically thin accretion disk, this radiation pressure instability manifests as a ``blown up'' solution, where the density is roughly constant over a large height of the disk structure, but it may happen that the density is larger around the photosphere than in the disk mid-plane.
In non-magnetic disk, this would satisfy the criterion for the convective instability and the density gradient would be restored \citep{1999-Rozanska}.

In our models, we do not include convection, as it might be restricted by the magnetic field, which shapes the disk structure.
Instead, we compute of the pressure gradient, $\propto \frac{d \ln \rho}{d \ln z}$, and check its minimal value.
The negative value means that the density inversion occurs and the model might not be stable.
We mark these models { clearly while presenting results of our computations. 
Nevertheless, since we do not solve time-dependent radial equations 
\citep{szuszkiewicz}, we are not able
to compare changes caused by TI to those generated by 
radiation pressure instability front, when we expect that the earlier 
moves in vertical, while the later - in radial direction. 
We plan to address this analysis in the future work.}


\section{Results}
\label{sec:res}

For better visualization of the physical conditions occurring on the border of disk and warm corona we are plotting  several quantities describing the processes which take place in magnetically supported and radiatively cooled accretion disks. From X-ray spectral fitting we measure an averaged temperature of the warm corona cooled by Comptonization:
\begin{equation}
\label{eq:tavg}
T_{\rm avg} = \tau^{-1} \int_0^{\tau_{\rm cor}} T d\tau\ .
\end{equation} 
Note, that for the purpose of this paper we adopt $\tau$ to be 
Thomson optical depth measured from the surface towards disk mid-plane.
This is due to the fact that electron scattering optical depth is commonly measured
during data analysis, when soft X-ray excess is fitted by Comptonization model. 
We define the base of the corona $\tau_{\rm cor}$ as $\tau$ for which the 
temperature reaches minimum. At some figures displaying vertical structure, we also mark the position of photosphere, where total optical depth: $\tau + \tau_{\rm ff}^{\rm P} = 1$ or 
the thermalization zone, where effective optical depth $\tau^* = \sqrt{\tau_{\rm ff}^{\rm P}(\tau + \tau_{\rm ff}^{\rm P})} = 1$, where $\tau_{\rm ff}^{\rm P}$ is the mean Planck opacity for free-free absorption. 

\begin{figure*}
    \centering
    \resizebox{\hsize}{!}{\includegraphics{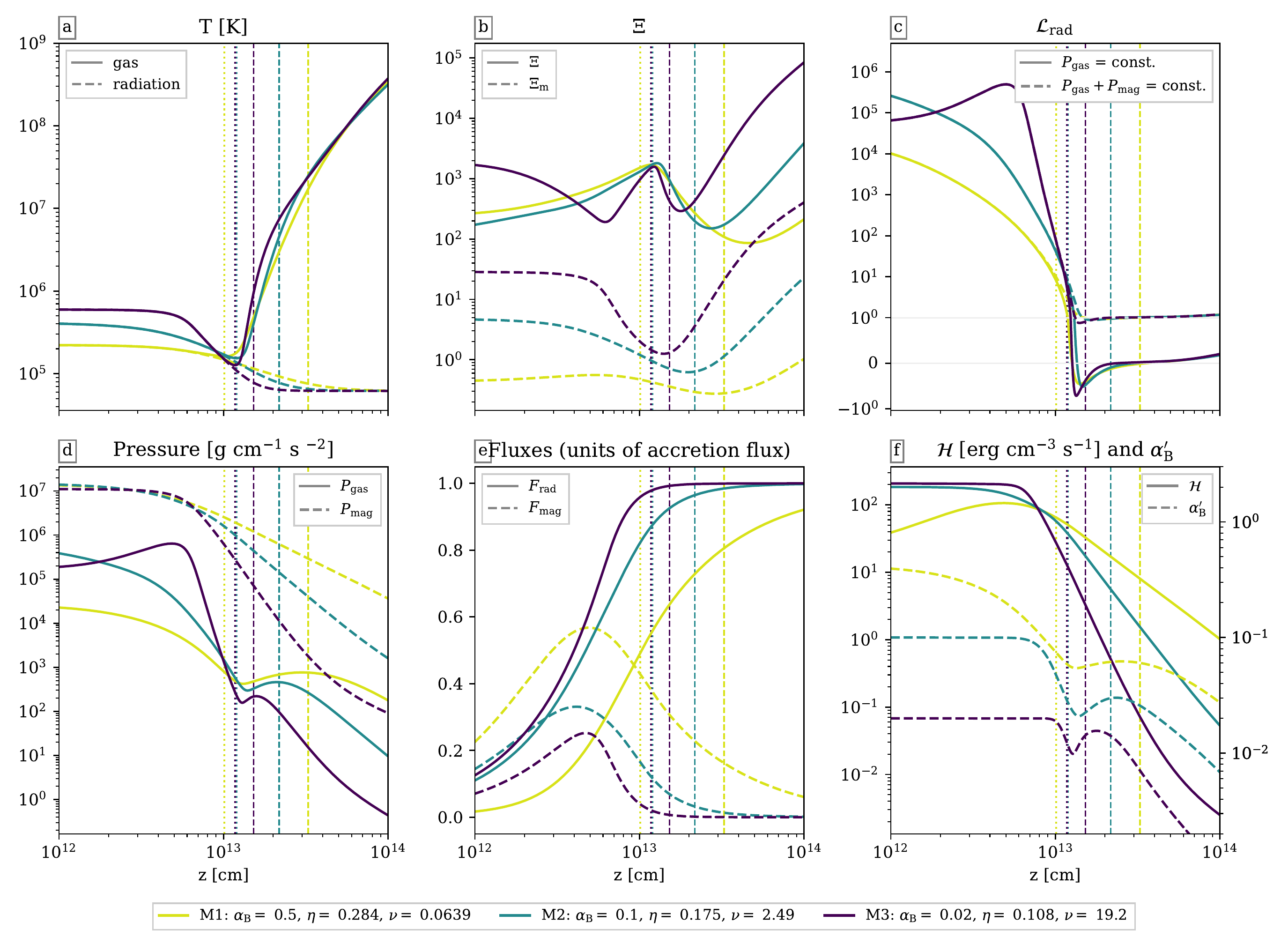}}
    \caption{Vertical structure of an accretion disk for our CM model versus $z$ in cm measured from disk mid-plane ($z=0$) up to the surface (right side of the figure panels),
     and  for three different sets of magnetic parameters ranging from the strongest (M1 - by yellow line) to the weakest (M3 - by black line) magnetic field, and listed in the box below the figure. Local temperature, ionization parameter (Eq.~\ref{eq:xi} and~\ref{eq:xim}), and stability parameter (Eq.~\ref{eq:instabil}) are plotted in the top row, while pressures, fluxes relative to the total dissipated flux, and heating rate are plotted in the bottom row, respectively. 
     In addition, the value of $\alpha_{\rm B}'$ according to Eq.~\ref{key} is shown in the last panel of second row, by dashed lines.
    The positions of corona base $\tau_{\rm cor}$ are marked by vertical dotted lines, and 
    the positions of photosphere - by dashed lines for each model, respectively. 
    }
    \label{fig:cute}
\end{figure*}

\begin{figure*}[t]
  \centering
  \resizebox{\textwidth}{!}
            {\includegraphics{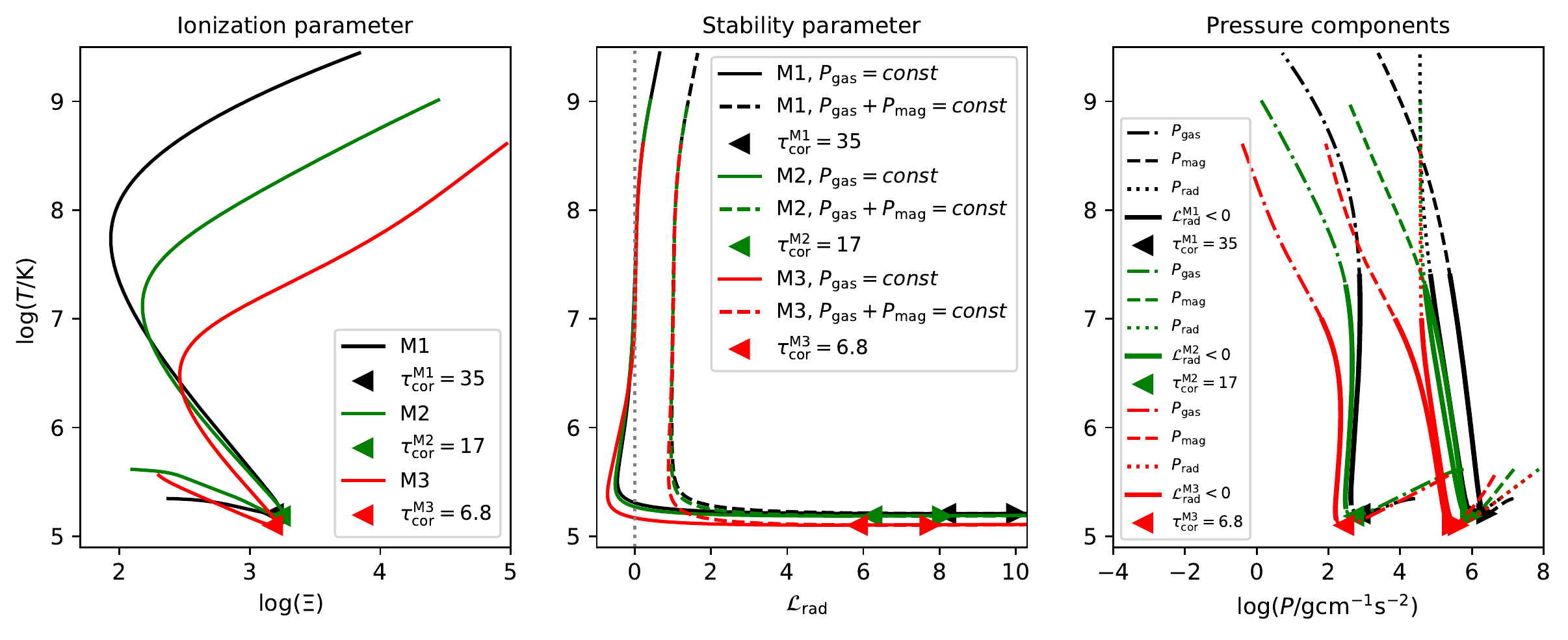}}
            \caption{ Temperature structure of an accretion disk for our CM model versus ionization parameter (left panel), stability parameter (middle panel) and all three pressure components (right panel). Gas pressure structure is given by dashed-dotted line, magnetic pressure by dashed line, and radiation pressure by dotted line. Three models of different magnetization are defined as in
 Fig.~\ref{fig:cute}, and given by colors: M1 -- black, M2 -- green, and M2 -- red. At each panel, the position of temperature minimum is clearly indicated by a triangle of the same color. Gray dotted line in the middle panel marks the limit of negative value of stability parameter. Thick solid lines in the right panel cover values of pressure and temperature for which classical stability parameter (under constant gas pressure) is negative.}
\label{moj:plot}
\end{figure*}

To trace TI the important quantity is well known ionization parameter \citep{krolik81,2015-Adhikari}, defined as: 
\begin{equation}
\Xi= \frac{P_{\rm rad}}{P_{\rm gas}}. 
\label{eq:xi}
\end{equation}
Although we do not compute ionization states of the matter, the above parameter indicates the place in the disk structure where TI physically starts \citep{1996-RozanskaCzerny,1999-Rozanska-Conduction}. The TI appears exactly when the gradient of cooling function, so called stability parameter computed under constant gas pressure becomes negative
\begin{equation}
\label{eq:instabil}
\mathcal{L}_{\rm rad} \equiv \left. \frac{d \ln \Lambda}{d \ln T} \right|_{\delta P_{\rm gas} = 0}  < 0 \ ,
\end{equation}
as we also visualize in our figures below. Our computations allow us to indicate Thompson optical depth, $\tau_{\rm min}$,  at which the above stability parameter is minimal.
{ In addition, we calculate the value of net cooling rate gradient over temperature 
under constant gas plus magnetic pressure in order to show how magnetic pressure affects 
stability of the disk heated by MRI}.  

To estimate the importance of the magnetic pressure versus gas pressure at each depth of  the disk, the commonly used magnetic pressure parameter defined as: 
\begin{equation}
\beta= \frac{P_{\rm gas}}{P_{\rm mag}}. 
\end{equation}
For the purpose of this paper in analogy do the ionization parameter we define the magnetic ionization parameter as:
\begin{equation}
\Xi_{\rm m}= \frac{P_{\rm rad}}{P_{\rm mag}+P_{\rm gas}}, 
\label{eq:xim}
\end{equation}
which helps us to show the domination of radiation pressure across the disk vertical structure.  
Furthermore, we have demonstrated in GR20 that the density of the corona depends mostly on the magnetic field gradient: 
\begin{equation} 
\label{eq:qcor}
    q = -\frac{d \ln P_{\rm mag}}{d \ln z}.
\end{equation}
Note, that all parameters change with disk/corona vertical structure and we mark them with ``0'', when defined at mid-plane. { We also define the magnetic dissipation rate as follows}:
\begin{equation}
\label{eq:rate}
q_{\rm h} =\left( \frac{m_{\rm H}}{ \rho}\right)^{2} {\cal H},
\end{equation}
in erg~cm$^3$~s$^{-1}$, so it can be compared to other heating/cooling rates
which are delivered in the literature in the same units. 
As an outcome of our computations we also deliver information about total surface density $\Sigma$  
in g~cm$^{-2}$ of the disk/corona system, which 
is the density integrated over total height.  

Finally, properties of the Compton cooled surface zone are observationally tested 
with the use of Compton parameter: 
\begin{equation}\label{eq:yavg}
y_{\rm avg} = \int_0^{\tau_{\rm cor}} \frac{4 k (T - T_{\rm rad})}{m_{\rm e} c^2} \left( 1 + 2 \tau \right) d\tau\ ,
\end{equation}
and in the section below we determine such parameter either up to thermalization 
layer $\tau^\star$ or up to the corona base $\tau_{\rm cor}$.

 In analogy to \cite{1991-HaardtMaraschi}, we denote the fraction of the energy released
 in the corona versus total thermal energy $F_{\rm rad}$ as
\begin{equation}\label{eq:chipar}
f = \frac{F_{\rm rad}^{\rm cor}}{F_{\rm rad}} = 1 - \frac{F_{\rm rad}^{\rm disk}}{F_{\rm rad}},
\end{equation}
$f = 0$ corresponds to the passive corona whereas $f = 1$ corresponds to passive disk. In case of magnetically supported disk $f$ depends only on magnetic parameters, and both fluxes $F_{\rm rad}^{\rm cor}$ and $F_{\rm rad}^{\rm disk}$ are not assumed, but numerically calculated in our model. The division between disk and corona is taken at the layer when coronal temperature reaches minimum i.e. at $\tau_{\rm cor}$. All quantities listed in this section are self consistently derived as an outcome of our model. We present them in order to check the model correctness and to compare them to observations.


\subsection{Vertical structure}

 In order to understand the formation of warm corona,
the vertical structure of our CM model is shown in Fig.~\ref{fig:cute}. In all panels the  horizontal axis is the height  above the mid-plane of the disk in units of cm.
Each figure panel represents the structure of parameter given in the panel's title.  Three models for various magnetic parameters are given by different line 
colors described in the box above figure caption. The model called M1 stands 
for a highly magnetic disk { with value of} $\alpha_{\rm B}=0.5$, while the model M3 stands 
for low magnetic disk -- { with} $\alpha_{\rm B}=0.02$. 

The gas temperature of the disk (panel \texttt{a} solid lines), { in all three models, follows the trend with warmer center and cooler photosphere which eventually undergoes an inversion and forms a corona being much hotter 
than the center of the disk.} 
The inversion occurs roughly at $z=10^{13}$\,cm.
The temperature of the corona rises sharply but gradually with height as $z^2$, even though the heating rate per volume, plotted by solid lines at the panel \texttt{f} of Fig.~\ref{fig:cute}, decreases as $z^{-q}$.
{ Closer to the equator, the temperature rise typically present in case of 
standards viscous $\alpha$-disk \citep{1973-ShakuraSunyaev} saturates 
 due to magnetic heating being distributed much further away from the equatorial plane, resulting
in lower radiative flux and less steep temperature gradient.}

\begin{figure*}
    \centering
    \resizebox{\hsize}{!}{\includegraphics{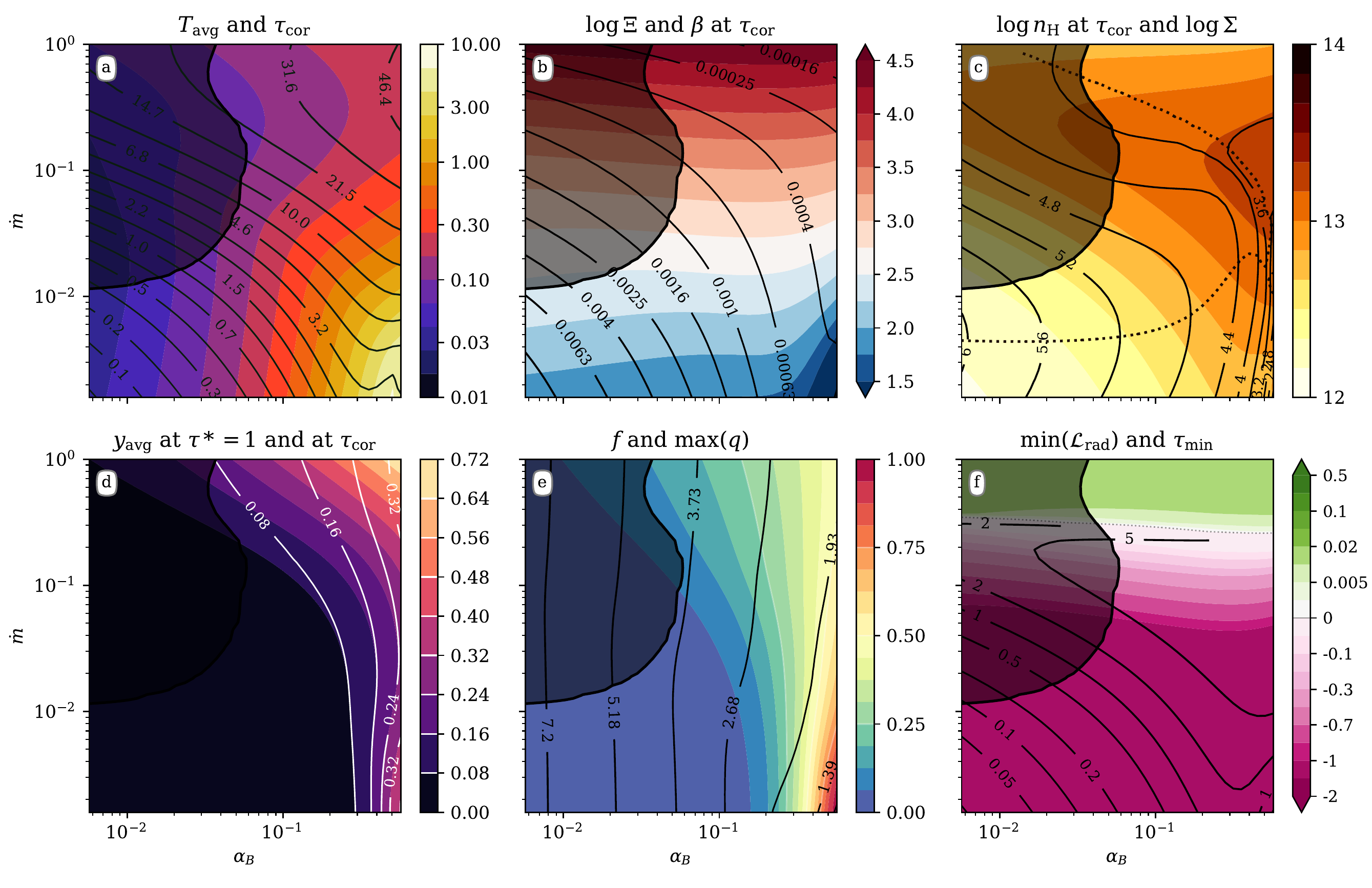}}
    \caption{The properties of the warm corona on accretion rate -- 
    magnetic viscosity parameter plane.  All models have been calculated for 
    $M_{\rm BH} = 10^8 M_\odot$ and $R  = 6 \, R_{\rm Schw}$.   
    Upper panels show: \texttt{a} - average temperature $T_{\rm avg}$ of the corona  in keV 
   ({ colors}) and optical depth of the base of warm corona $\tau_{\rm cor}$ ({ contours});  \texttt{b} - $\Xi$ ({ colors}) and 
   magnetic pressure parameter $\beta$ ({ contours}) both at  $\tau_{\rm cor}$, 
   and \texttt{c} - number density $n_{\rm H}$ in cm$^{-3}$ ({ colors})
    at $\tau_{\rm cor}$ and total column density $\Sigma$ in g~cm$^{-2}$ ({ contours}). 
    Dotted contour here shows where $d \Sigma / d \dot m = 0$.
    Bottom panels display: \texttt{d} -  $y_{\rm avg}$ parameter of the warm corona 
    at the thermalization zone i.e. $\tau^*=1$ ({ colors}) 
    and at the base of corona $\tau_{\rm cor}$ ({ contours}), \texttt{e} - 
    fraction of radiative energy produced by the corona $f$ ({ colors}) 
    and the maximum value of magnetic field gradient $q$ ({ contours});
    and \texttt{f} - 
        the minimum value of $\mathcal{L}_{\rm rad}$  throughout the disk height ({ colors})
        and Thomson optical depth $\tau_{\rm min}$ when it happens ({ contours}),
         green areas are stable disks, whereas magenta areas indicate thermal instability in the corona, the gray dotted line corresponds to $\mathcal{L}_{\rm rad} = 0$.
        In all panels, dark contour indicates the parameter sub-space affected by the density inversion.
    }
    \label{fig:maps-MA}
\end{figure*}

The radiation pressure is large in the core, between 2-3 orders of magnitude higher than 
the gas pressure, and  can be of comparable magnitude to the magnetic plus gas pressure
(panel \texttt{b}). Whereas it fully dominates at the atmosphere where warm corona is formed, where both ionization parameters, $\Xi$ and $\Xi_{\rm m}$ increase by two orders of magnitude with the height towards the surface. In addition, the inversion of both parameters is present, regardless of the value of magnetic pressure. 
Such a course of $\Xi$ and $\Xi_{\rm m}$  is connected with the structure of stability parameter, presented in panel \texttt{c} of Fig.~\ref{fig:cute},  which is negative 
even for the most stringent condition of constant gas pressure. Although not large geometrically, this area actually constitutes a large part of the optically thick corona.

The gas pressure (panel \texttt{d}) is roughly constant in the center  (due to radiation pressure) and decreases as $z^{-(q+2)}$ with exception of the model M3 which has the weakest magnetic field.
In that model, the region of lesser gas pressure forms in the core of the disk, due to radiation pressure being a primary support for the disk structure, which is because convection that would remove this inversion is absent in our model.
In the region of the photosphere the gas pressure levels out, just to enter a sharp decrease in the magnetic pressure dominated corona.
It is worth noting that the magnetic pressure is dominating significantly over gas pressure in our model at all points of the vertical structure.

The radiative and magnetic fluxes are released at a comparable rate inside the disk (panel \texttt{e}),  where the fluxes are given in relation to the total flux released by accretion.
In the magnetic dominated corona, the magnetic flux is quickly depleted in sub-inversion area, and it is almost zero when it enters the corona  with the exception for strongly magnetized models.

The magnetic heating in the corona is decreasing sharply  with height
(panel \texttt{f}), except for the strongest magnetic field case M1, 
where the peak heating occurs in the transition region and then 
${\cal H}$ gently decreases.
In this model, the MRI process begins to quench at around $z=10^{12}$cm, 
which is ten times lower than other models, and can be seen as a decrease 
in $\alpha_{\rm B}'$, (given by the dashed lines at the same panel) 
slightly above equatorial plane.
By $z=3 \times 10^{13}$cm, the MRI becomes almost completely inactive, 
due to rise in the gas pressure in the corona. To produce such corona, 
the value of toroidal magnetic field at the equatorial plane is 
$\log B_{0} = 4.25$, 4.28, and 4.22 G, for considered models: M1, M2 and M3,  
respectively, and it is three orders of magnitude lower than 
in case of GBHB (GR20).

{ To follow the vertical extension of TI zone, we present the temperature 
structure versus ionization parameter defined by Eq.~\ref{eq:xi} on 
Fig.~\ref{moj:plot} left panel, for our three models M1, M2, and M3 
of different magnetization. In case of each model, we mark the corona 
base $\tau_{\rm cor}$ by a triangle. The part of stability curve with 
negative slope is clearly present in each of the model, which clearly 
shows that magnetic heating does not remove TI  in regions of high 
temperatures, of the order of 1 keV, and high densities, of the order 
of $10^{12}$\,cm$^{-3}$, cooled by radiative processes. Interestingly, 
such TI is created only for Compton and free-free absorption/emission. 
Going down with temperature, below the corona base, there is another 
turning point. 
Below this point, the gas stabilizes due to magnetic heating. It can be 
firstly deduced from the fact that going towards the equatorial plane the 
disk temperature increases, which is clearly presented in panel a 
of Fig.~\ref{fig:cute}.}

{ In order to show how magnetic pressure affects the warm heated layer 
above cold disk, we plot stability parameter on Fig.~\ref{moj:plot} 
middle panel, for all three models, in two versions. Solid line presents 
classical cooling rate gradient under constant gas pressure, while dashed 
line the same gradient under constant gas plus magnetic pressure. It is 
clear, that negative value of the stability parameter occurs only in 
the case of classical assumption of constant gas pressure. Magnetic 
pressure directly makes cooling rate gradient to be always positive, 
acting as a freezer for eventual evolution of gas due to TI. For all three models, 
the base of the corona occurs below TI zones under the classical constant gas pressure condition.
 In the forthcoming paper, we 
plan to calculate proper timescales for TI evolution and for the existence of magnetically 
heated corona.}

{ In magnetically supported disk, the radiation pressure dominates in 
the disk interior, while it becomes comparable to magnetic pressure 
in thermally unstable zones. 
It is clearly demonstrated on the right panel of Fig.~\ref{moj:plot}, 
where solid lines cover pressure plots for regions where classical 
stability parameter is negative. Gas pressure, which reflects the 
gas structure, is always at least four orders of magnitude lower 
than magnetic pressure. It is worth to note, that going from the 
surface towards corona base, the gas temperature decreases while 
all three pressure components tend to increase. After passing 
$\tau_{\rm cor}$, the temperature starts to increase with 
depth and with the increase of all three pressure components. 
Small negative slopes 
in the vertical structure of gas pressure do not affect the overall 
stability of the disk, since the gas distribution is mostly kept by 
magnetic pressure.}

\subsection{Warm corona across the parameter space}
\label{warm:par}

\begin{figure*}
    \centering
    \resizebox{\hsize}{!}{\includegraphics{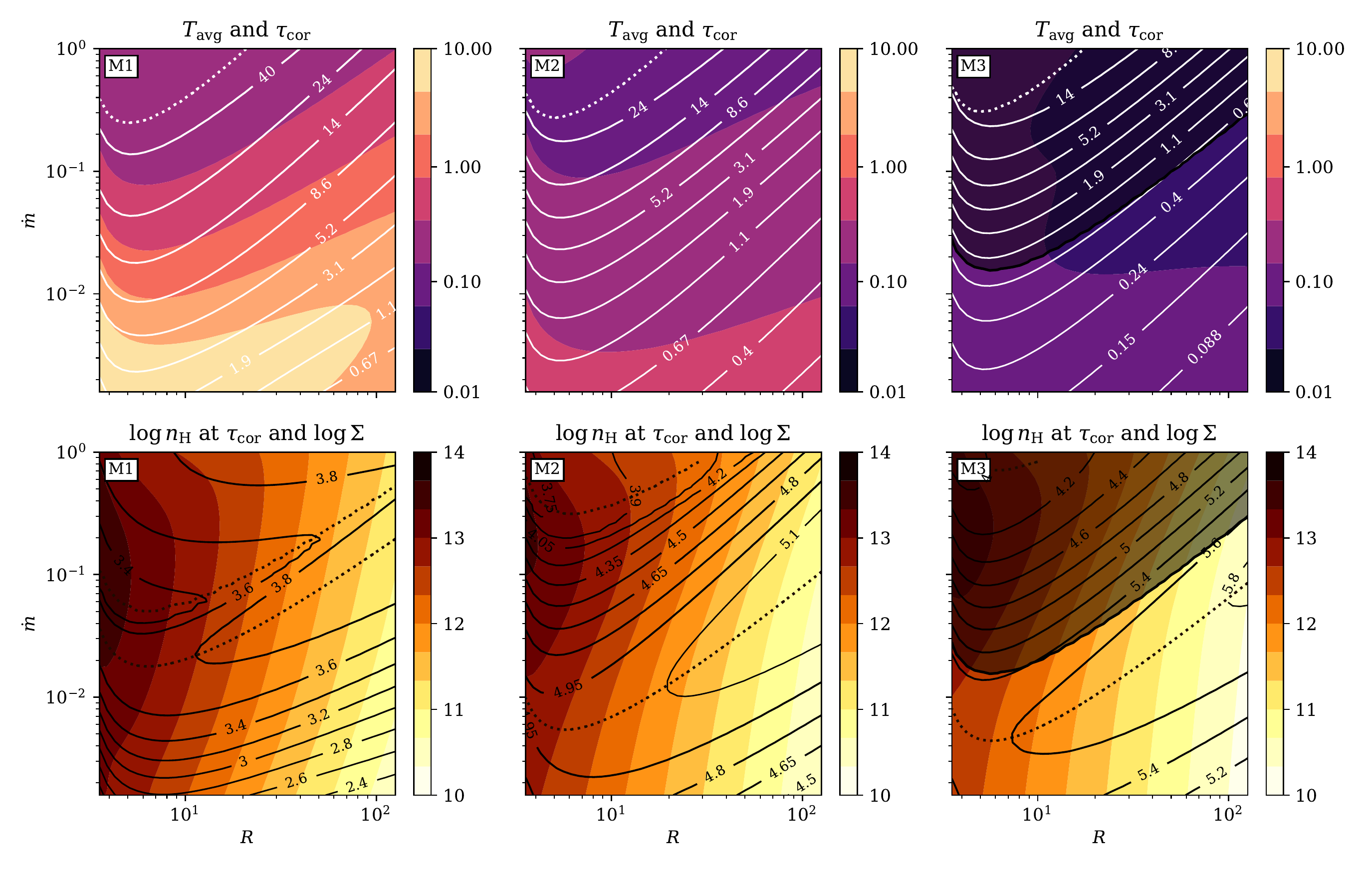}}
    \caption{The properties of the warm corona on accretion rate -- radius plane given in 
    units of $R_{\rm Schw}$.
        All models have been calculated for $M_{\rm BH} = 10^8 M_\odot$, and magnetic
        parameters are corresponding M1-  left, M2 - middle, and M3 - right panel columns
        with values given under Fig.~\ref{fig:cute}.  
        Upper panels display average temperature $T_{\rm avg}$ of the corona  in keV 
   (colors) and optical depth of the base of warm corona $\tau_{\rm cor}$ (contours).
        Dotted contours at upper panels indicate where $\mathcal{L}_{\rm rad} = 0$
        under constant gas pressure.
    Bottom panels show number density $n_{\rm H}$ in cm$^{-3}$ (colors)
    at $\tau_{\rm cor}$ and total column density $\Sigma$ in g~cm$^{-2}$ (contours). 
     Dotted contours at bottom panels show where $d \Sigma / d \dot m = 0$.
     Dark contours in case of M3 model, indicate the parameter sub-space affected
     by the density inversion.
    }
    \label{fig:map-spec-R}
\end{figure*}

The dependence of coronal parameters derived from our model on the accretion rate and on the magnetic field strength for black hole mass 
$M_{\rm BH} = 10^8 M_\odot$ and at the radius $R  = 6 \, R_{\rm Schw}$
is shown in Fig.~\ref{fig:maps-MA}. 
We  adopt that the corona extends from the zone where the gas temperature reaches minimum, as visible at panel \texttt{a} of Fig.~\ref{fig:cute}, up to the top of an atmosphere.
The optical depth, $\tau_{\rm cor}$, is greater for higher accretion rate and stronger magnetic viscosity parameter. 
On the other hand, the temperature $T_{\rm avg}$, averaged over $\tau$ from the surface
down to $\tau_{\rm cor}$, 
 seems to be mostly dependent on the magnetic field strength.

As can be read from the ratios of radiation, gas and magnetic pressures, displayed in 
the panel \texttt{b} of Fig.~\ref{fig:maps-MA}, by plotting $\Xi$ (color maps) and $\beta$ (contours) at the corona base, the disk is dominated by the radiation pressure for the whole parameter space. In addition, magnetic pressure overwhelms  the gas pressure,
 and this is the main reason that we obtain optically thick soft corona layer as predicted by \citet{rozanska15}. The magnetic pressure is limited by the condition in Eq.~\ref{eq:pmagmax}, which occurs  as a sharp transition in all global parameters at around $\dot m > 0.01$. For such large accretion rate, the contribution of radiation pressure is very high, and if the magnetic support is not strong enough, the solution becomes unstable (see Sec.~\ref{sec:inflated}), which is marked as a black overlay in all panels.
This effectively restricts the solutions for high accretion rates only to models where the magnetic field is sufficiently strong ($\alpha_{\rm B} > 0.05$).

The gas density  at the base of corona (panel \texttt{c} of Fig.~\ref{fig:maps-MA}) is the highest for both  high  accretion rate and high magnetic parameter.
 Decreasing the magnetic field also has a global stability effect on the disk.
For a thin accretion disk, the quantity $d \Sigma / d \dot m$ should be positive, otherwise we encounter a problem with well-known stability of standard viscous $\alpha$-disk \citep{1973-ShakuraSunyaev}. 
For a weak magnetic field, this value is negative even for $\dot m = 0.003$, as indicated by the region enclosed by the dotted contour in panel \texttt{c} of Fig.~\ref{fig:maps-MA}.
If we introduce a stronger magnetic field, however, the band of accretion rate where this criterion is met, is narrowed down considerably.

Investigating the $y_{\rm avg}$ parameter (panel \texttt{d} of Fig.~\ref{fig:maps-MA}) and comparing it to corona-to-total flux ratio $f$ (panel \texttt{e} of Fig.~\ref{fig:maps-MA}) we find similar behavior, where the largest values of both parameters are obtained for  
$\alpha_{\rm B} > 0.3$.   Nevertheless, the Compton parameter has his largest values for high accretion rate, while the amount of energy generated in the corona is largest for low values of accretion rate.

The magnetic field gradient parameter $q$  is mostly imposed by the model parameters, but it is worth seeing that its value only drops below 2 for the most magnetized models
(panel \texttt{e}).
On the other hand, the value of stability parameter $\mathcal{L}_{\rm rad}$ presented
in panel \texttt{f} is almost exclusively dependent on the accretion rate, while the optical depth $\tau_{\rm min}$, can reach values up to 5.

\begin{figure*}
    \centering
    \resizebox{0.8\hsize}{!}{\includegraphics{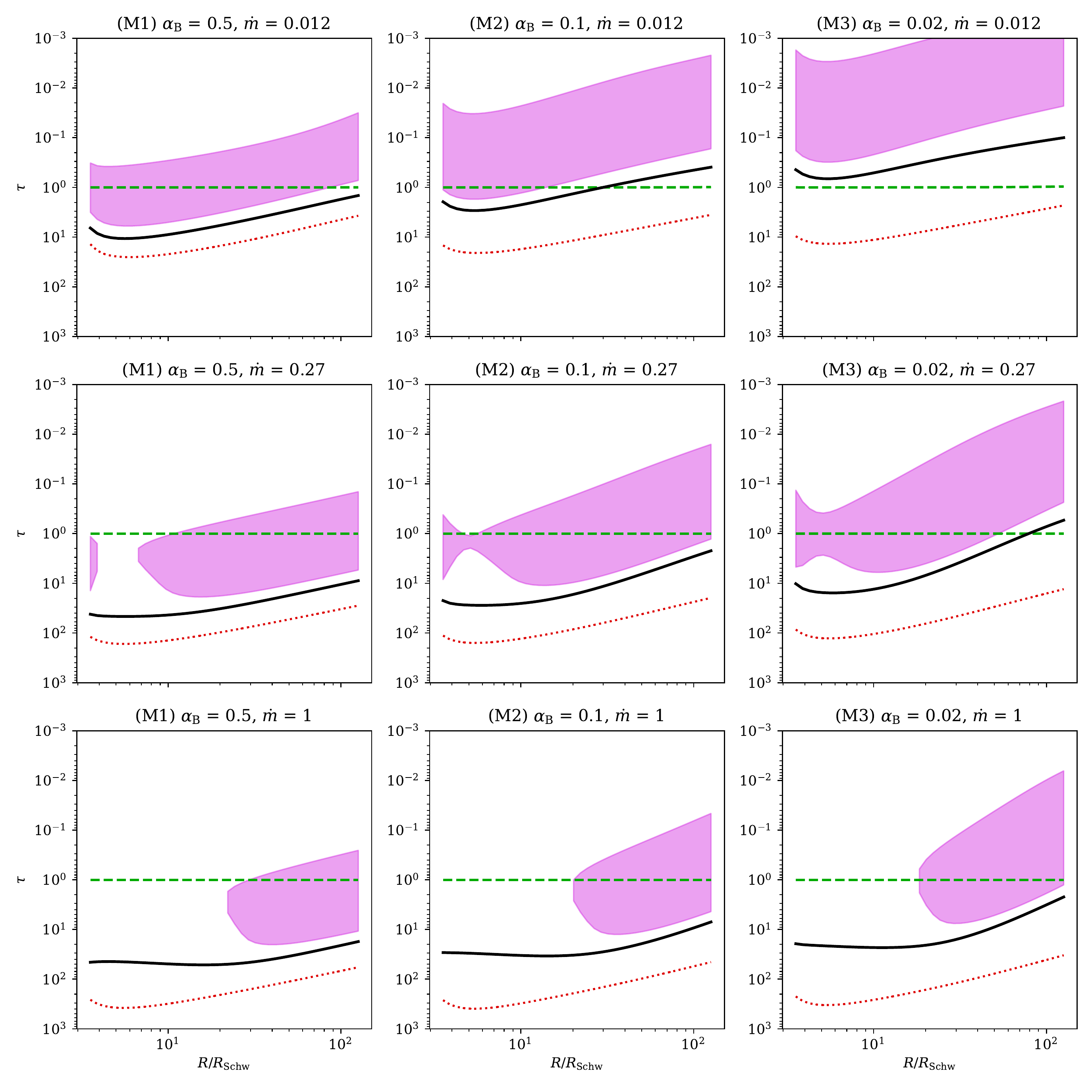}}
    \caption{Thomson scattering optical depth of TI zones, plotted 
    against the radial distance from the BH of the mass $M_{\rm BH} = 10^8 M_\odot$, and shown for three values
    of magnetic parameters M1, M2, M3 (from left to right columns) and accretion rate $\dot m = 0.12, 0.27$ and 0.1 (from upper to bottom rows) .
        Photosphere is marked using dashed green line, temperature minimum - 
    using thick black solid line, thermalization depth ($\tau*=1$) using red dotted
    line. The extent of thermal instability is presented as pink filled area.}
    \label{fig:radial_tau}
\end{figure*}
 
\subsection{Radial structure of the warm corona}

Fig.~\ref{fig:map-spec-R} shows the radial dependence of global parameters of warm corona for our canonical BH mass 
$M_{\rm BH} = 10^8 M_\odot$, and for three strengths of the magnetic field, i.e. M1, M2, and M3 models, with the same magnetic parameters $\alpha_{\rm B}$, $\eta$, 
$\nu$ as in Fig.~\ref{fig:cute}. Upper panels show maps of $T_{\rm avg}$ given by Eq.~\ref{eq:tavg},  and contours of  $\tau_{\rm cor}$.
Dotted line contour separates thermally stable regions for high accretion rates
and thermally unstable for low accretion rates by the condition
$\mathcal{L}_{\rm rad} = 0$.
Bottom panels present maps of the density number of the warm corona and contours 
indicate the total column density $\Sigma$ computed form the top down to the mid-plane. 
In case of M3 model, with the lowest magnetization, we clearly indicate sub-space affected by classical radiation pressure instability marked by dark contours.

\begin{figure*}
    \centering
    \resizebox{\hsize}{!}{\includegraphics{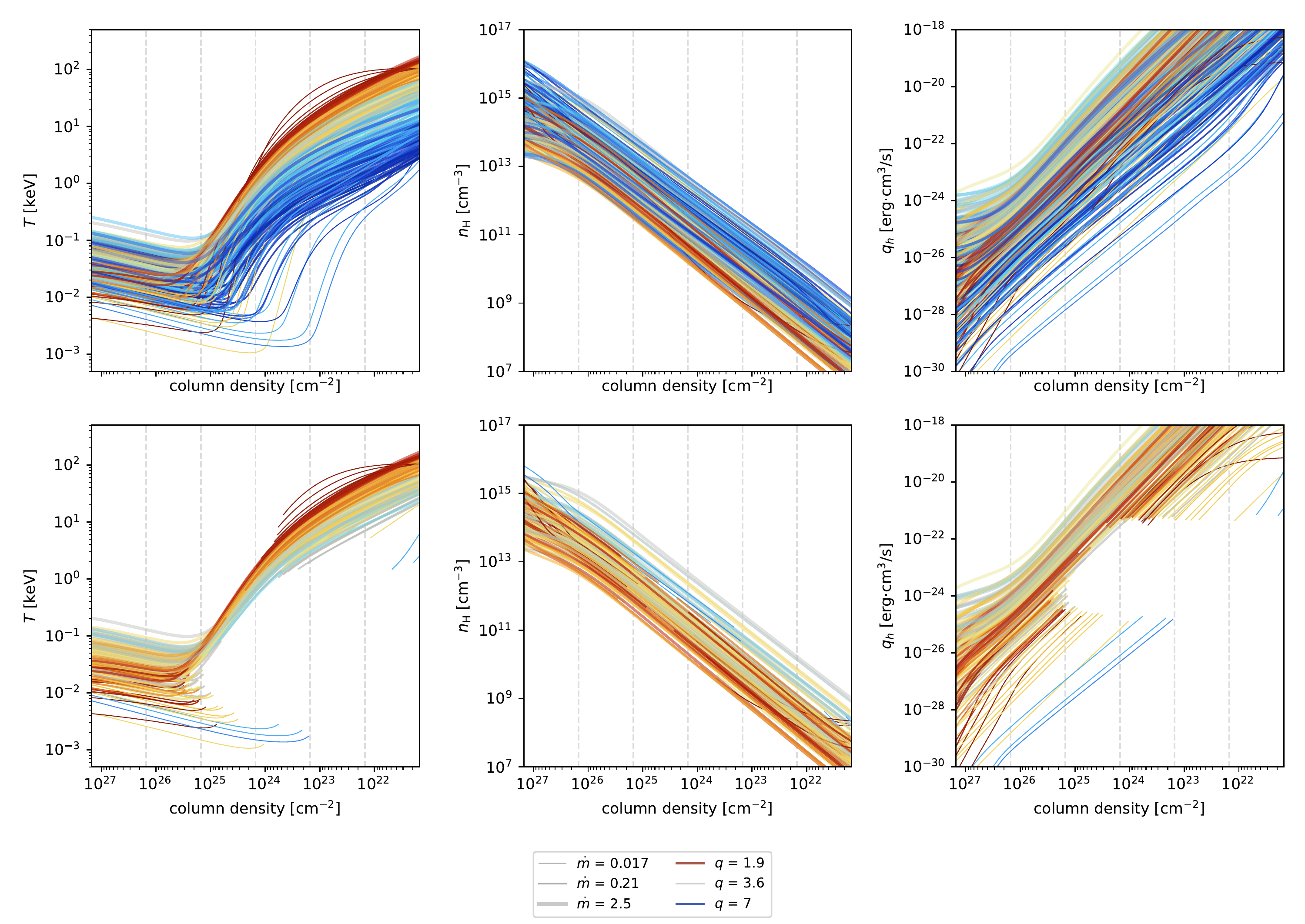}}
    \caption{Vertical structure of temperature, density and dissipation heating rate 
    defined by Eq.~\ref{eq:rate} for sample of models computed according to 
    the method described in Sec.~\ref{sec:ran}. 
    The first row shows the complete sample of models, while the second row only shows models that do not exhibit density inversion discussed in Sec.~\ref{sec:inflated}.
    Zones  where TI occurs are omitted, which produces the gaps visible
    in many models. 
   The color of the line corresponds to the magnetic gradient $q$ (red is stronger field, blue is weaker), while the thickness shows the accretion rate $\dot m$ (thicker is higher). The gray vertical lines show the optical depths $\tau$ equal to (from left to right): 100, 10, 1, 0.1, 0.01. 
    }
    \label{fig:cfxy-6}
\end{figure*}

The overall radial dependence of the warm corona parameters follows the trend that corona is hotter and more optically thick for strongly magnetized disk 
(M1 model on left panels). Higher accretion rate makes corona cooler, i.e. lower $T_{\rm avg}$, but optically thicker. Such warm and optically thick corona in case of high accretion rate is rather more dense then in case of hotter corona appeared for low accretion rate.
However, for the model with a strong field, the quantity $d \Sigma / d \dot m$, which is associated with stability of viscous accretion disk, is positive in all but a narrow stripe in the parameter space marked by dotted lines at bottom panels of Fig.~\ref{fig:map-spec-R}.
On the other hand, for a weak field, M3 model, there is a maximum column density for around $\dot m = 0.03$ above which $\Sigma$ decreases.

In order to specify how TI, caused by gas cooling processes, affects the existence of warm corona we plotted 
the vertical extent of thermally unstable zones versus distance from the black hole 
in Fig.~\ref{fig:radial_tau} for three different magnetic parameters and three accretion rates.
The extend of TI zone is presented as pink filled area. 
The base of the corona is clearly marked by thick black solid line. 
The optical depth of the corona base is larger 
for higher accretion rate.
{ As it was shown in the previous section, TI indicated by negative values of the stability parameter is present for the broad range of parameters under the assumption of constant gas pressure, but disappears when it is computed over constant sum of gas and magnetic pressure.
}
Furthermore, it is produced by radiative processes, as Comptonization and free-free emission/absorption, being the only cooling processes in the models considered in this paper. Ionization/recombination may influence the extend of TI \citep{1996-RozanskaCzerny}, 
but the final result cannot be analytically predicted. Further computations are needed to show 
whereas ionization/recombination decreases or increases thermally unstable zone. 
Only for high values of accretion rate, higher than $0.2$, { and only in the inner regions of an 
accretion disk within 20~$R_{\rm Schw}$,
the existence of 
thermally stable warm corona with classical condition of constant gas pressure, 
is possible. 
We claim here, that in case of MSD, the condition of thermal stability under constant total i.e. gas plus magnetic pressure should be used in order to estimate if the TI zone is frozen into magnetic field. In the forthcoming paper, we plan to calculate relevant timescales for this kind of radiative cooling in the magnetically supported plasma. 
}


\subsection{Random models sample}
\label{sec:ran}

Our numerical code (GR20) allows us to calculate  a huge number of 
models in a finite time.
In order to put constrains on basic observational parameters resulting from our model: temperature of the corona and its optical depth, we
 have selected results from random distribution of six input parameters for our calculations described in Sec.~\ref{sec:params}, to account for their uncertainty.
We use the notation where $U(a,b)$ is a random variable with uniform distribution between $a$ and $b$ and $N(\mu,\sigma)$ is a normal distribution with expected value $\mu$ and spread $\sigma$.
We use a fixed radius $R = 6 R_{\rm Schw}$, while black hole mass $M_{\rm BH}$ and
accretion rate $\dot M$ (in g\,s$^{-1}$) are selected in the way to mimic the 
distribution of the same quantities in the sample of 51 AGN 
for which the warm corona was observed \citep[][see section below]{jin12}.  
The random walk overdoes through the following distribution:
\begin{equation}
\label{eq:bhm}
 \log \left( M_{\rm BH} \right)  = N(7.83, 0.63)\, ,
\end{equation}
\begin{equation}
\label{eq:acr}
\log ( \dot M )  = 0.27 \left( \log \left( M_{\rm BH}  \right) - 8 \right) + N(25.83, 0.52) \, ,
\end{equation}
where the expected and spread values were estimated from the observed sample.
 After running the code and solving the structure, we reject the models that satisfy the radiation pressure criteria described in Sec.~\ref{sec:inflated}.

We generated a set of models based on random distribution of the 
global disk parameters described by Eqs:~\ref{eq:bhm} and \ref{eq:acr}.
The vertical structures of all models in the sample are plotted collectively in Fig.~\ref{fig:cfxy-6}, the full sample (upper panels) as well as sample clipped according to  the criterion of the density inversion (lower panels) described in Sec.~\ref{sec:inflated}.
The temperature inversion occurs between $\tau=0.01$ and $\tau=30$ for all models, with both magnetic field and accretion rate having a positive effect on the corona depth, as shown in left panels of both rows of Fig.~\ref{fig:cfxy-6}.
The density follows a rather predictable relation $n_{\rm H} \propto z^{-q-2} \propto \left(N_{\rm H}\right)^{\frac{q+2}{q+1}}$ and the heating rate changes as $q_{\rm h} \propto z^{q+4} \propto \left( N_{\rm H} \right)^{-\frac{q+4}{q+1}}$, with small fluctuations due to the opacity.

When we remove the models that do not satisfy our criterion for the density gradient, i.e. we switch from upper to lower panels of Fig.~\ref{fig:cfxy-6}, the picture does not change much, but it becomes more complex when we omit the parts of the structure which is thermally unstable (when $d \ln \Lambda / d \ln T < 0$).
For the set of models we analyzed, the TI is confined to the area where 
$10^{-24} \leq q_{\rm h} \leq 10^{-22}$, and this does not seem to strongly vary with any of the model parameters we consider. Thus, we can summarize, that independently on the model parameters, 
those associated with accretion disk or with magnetic field, { the classical TI (under constant 
gas pressure)} occurs for magnetically 
supported disk with radiative processes as Comptonization and free-free emission. 


\subsection{Observational predictions}
\label{obs:pred}

As the main outcome of our computations, we obtain values of warm corona temperatures 
and optical depth, which can be directly comparable to the measurements reported in the 
literature. The data analysis concerning warm corona require high resolution, sensitive telescopes to collect enough photons in the soft X-ray energy range around 1 keV. The overall modeling is time consuming and usually one paper is devoted to one source. We have 
looked at the literature and we chose those sources for which both parameters: warm corona temperature and its optical depth, were measured. 
The warm corona measurement for the sample of objects was presented only by 
\citet{jin12}, who collected 51 sources. We take all those points into account while comparing to data. In addition 
we selected measured warm corona parameters of individual sources found in
\citet{1998-Magdziarz,page2004,2011-Mehdipour-Mrk509,2012-Done-SXE,2013-Petrucci-Mrk509,2014-Matt,2015-Mehdipour,2018-Middei,2018-Porquet,2018-Ursini,2019-Middei}.
All data points compared to our model in Sec.~\ref{obs:pred} below,  are listed in Tab.~\ref{tab:obs}.

\begin{table}[]
\caption{Observational points from literature compared to our model.}
\begin{center}
\begin{tabular}{ l|l|l}
 Source & Type  & Ref. \\
 \hline 
 \hline 
NGC 5548 & Sy1   &  \citet{1998-Magdziarz}  \\
\hline 
Q 0056-363  & QSO  & \multirow{3}{*}{\cite{page2004} }   \\
Mrk 876 & QSO  &   \\ 
B2 1028+31 & QSO  & \\ 
\hline
Mrk 509 & Sy1.5  &   \citet{2011-Mehdipour-Mrk509} \\ 
\hline 
RE 1034+396  & NLS1  & \multirow{2}{*}{\citet{2012-Done-SXE}}  \\
PG 1048+213  & BLS1    & \\
\hline 
51 sources & AGN1   & \citet{jin12} \\
\hline 
Mrk 509 & Sy1.5 & \citet{2013-Petrucci-Mrk509} \\
\hline 
Ark 120 & Sy1  & \citet{2014-Matt}\\
 \hline 
NGC 5548 &  Sy1  & \citet{2015-Mehdipour} \\ 
\hline 
NGC 7469  & Sy1  & \citet{2018-Middei} \\
\hline 
Ark 120 & Sy1  & \citet{2018-Porquet} \\
\hline
3C 382 & BLRG  & \citet{2018-Ursini} \\
\hline 
NGC 4593 &Sy1   & \citet{2019-Middei} \\
\hline 
\end{tabular}
\label{tab:obs}
\end{center}
\end{table}

\begin{figure*}
    \centering
    \resizebox{0.95\hsize}{!}{\includegraphics{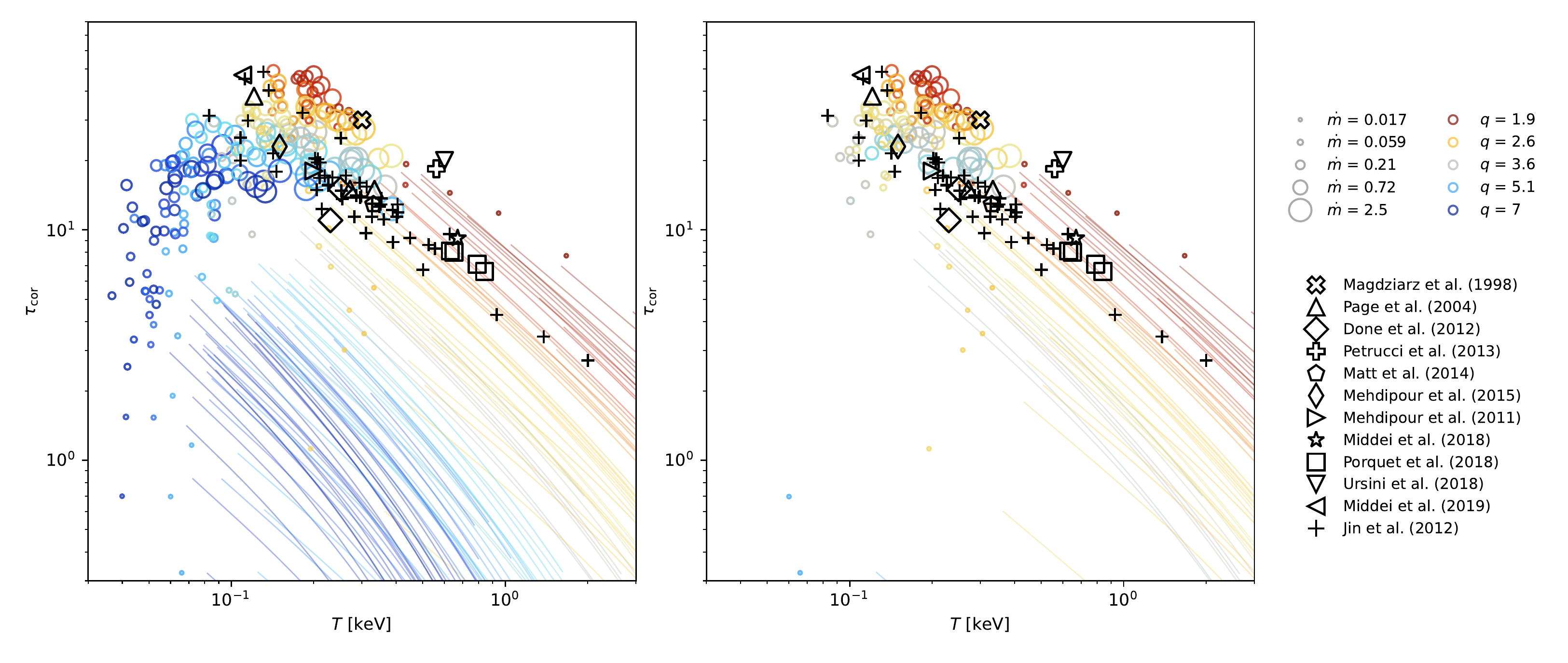}}
    \caption{
        Parameters of the corona $\tau_{\rm cor}$ and $T_{\rm avg}$ determined for a random sample of models are marked by open circles.
        The size  of the points determines the accretion rate and color of the points corresponds to the magnetic field gradient (similar to Fig.~\ref{fig:cfxy-6}) given on the 
      right side of the figure.
        For models where the TI is present, the values of $\tau$ and $T_{\rm avg}$ computed at every point of the instability are displayed by thin lines. 
        In the left panel, all models are shown, while in the right panel, models where the density inversion occurs (see Sec.~\ref{sec:inflated}) are filtered out. 
       Additionally, data points from the literature are drawn using black markers listed on the right side of the figure. The data points clearly follow TI strips.
    }
    \label{fig:obspoints1}
\end{figure*}

\begin{figure*}
    \centering
    \resizebox{0.8\hsize}{!}{\includegraphics{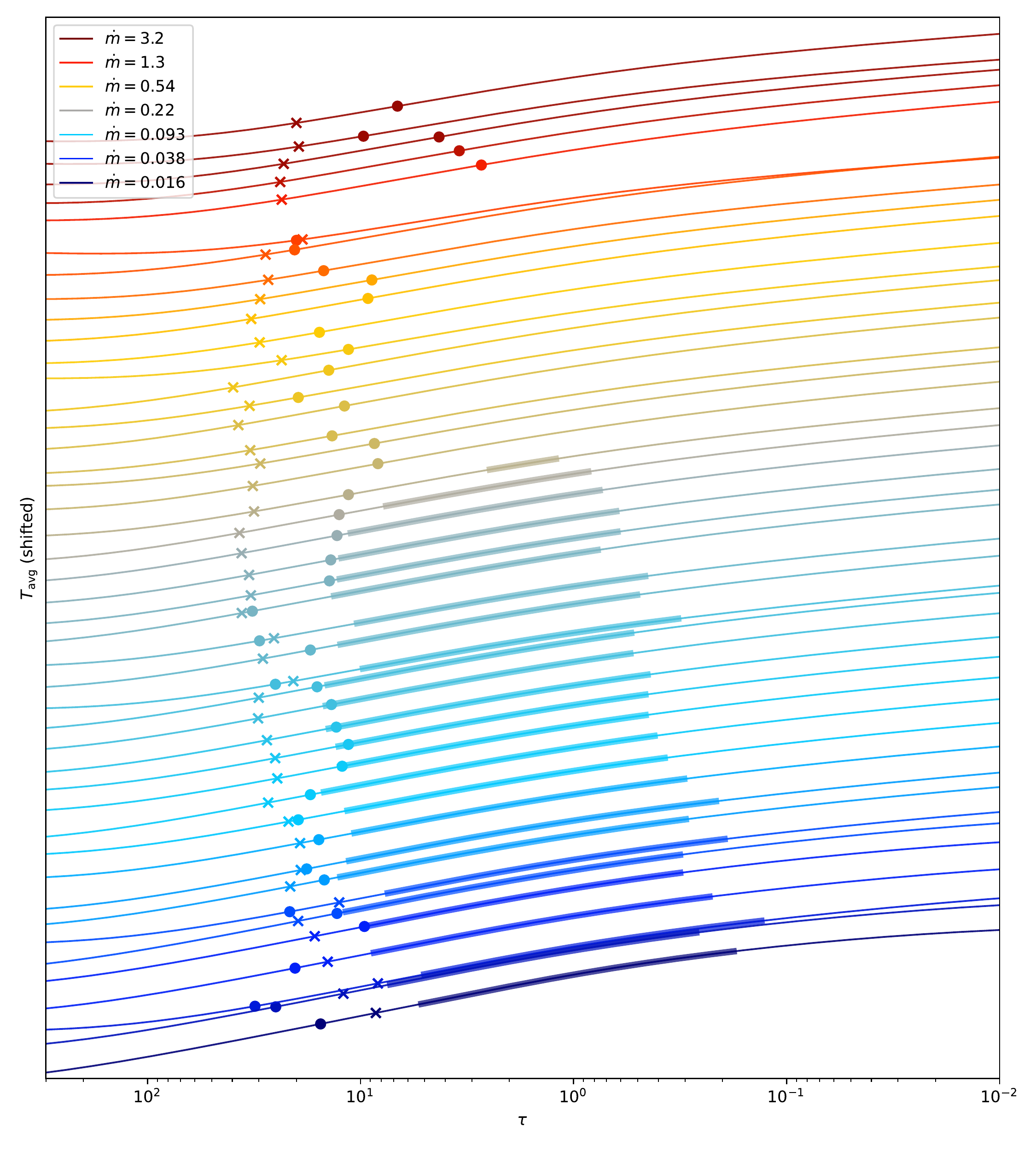}}
    \caption{Profiles of averaged temperature versus the optical depth for fine-tune models. 
    The temperature is averaged between the optical depth given by the coordinate of the horizontal axis and the surface of the corona.
    Thicker lines are overlayed to indicate where the condition for the TI is fulfilled.
    Circle markers are temperatures and optical depths obtained by \cite{jin12}.
    Cross markers indicate the averaged temperature at $\tau_{\rm cor}$ of a given model. 
    Magnetic parameters are adjusted so that the temperature profile 
    crosses the observational point.
    Each profile is shifted slightly for a clearer presentation, and models are sorted according to the accretion 
    rate $\dot m$.}
    \label{fig:jin_profiles}
\end{figure*}

From our random sample of models, we extracted essential parameters 
as optical depth of the warm corona, its temperature, accretion rate, 
black hole mass and magnetic field gradient, all of them 
described on the beginning of Sec.~\ref{sec:res}. 
{ The results are shown in Fig.~\ref{fig:obspoints1}, where the random 
values described in Sec.~\ref{sec:ran} are selected in the way 
to roughly match the observational points taken from Tab.~\ref{tab:obs}, in terms of accretion rate and the black hole mass.}
At the same times, we allowed the magnetic parameters of the disk to vary in a very broad range, from $\beta_0 \approx 10^3$ to $\beta_0 \approx 1$.
This allows us to scan all possibilities while remaining close to observations in terms of known quantities. Observations are marked on the right side of the figure according 
to data analyzed in papers listed in Tab.~\ref{tab:obs}. Some source are reported twice or even more, but it does not matter since we are interested in general observational trend of
the warm corona. 

In the left panel of Fig.~\ref{fig:obspoints1}, we include all results of our models by colored open circles, with  no limit to radiation pressure or viscous instability (as long as the model has converged).
The size of the circle indicates the value of accretion rate, while its color reflect the value of magnetic field gradient, both given on the right side of the figure. 
The cloud of points on the $\tau_{\rm cor}$--$T_{\rm avg}$ plane coincide with the observational results, but the range of corona parameters obtained
with our numerical code is broader. 
A clear correlation of magnetic field strength and the accretion rate with resulting corona parameters is seen. Observations follow the models computed for 
higher accretion rates and moderate magnetic field gradient.

In the right panel of Fig.~\ref{fig:obspoints1}, we removed all the models 
where the density inversion occurs, which are the models dominated by the radiation pressure and not stabilized by magnetic pressure. We reject these models because, unless convection is present to transfer the 
energy from the disk, the solution becomes ``puffed up'' and generally unstable. 
The models that are removed in this step are mostly models when $\tau_{\rm cor} <10 $ or optically thick models with high accretion and low magnetic field. 
The data points seem to group around the models which do not display TI in the structure of warm corona, but some of them follow thin lines, which are thermally unstable zones. The distribution among instability strips
can be attributed to our choice of the cutoff of the corona base at the temperature minimum. Had it been chosen according to a different criterion, a more shallow and hotter corona would be obtained. TI might cause the corona to be shallower than the temperature minimum, which could potentially happen at undetermined point in the vertical structure.

To check at which point of the vertical structure the data is in line with model,  we took the black hole mass and the accretion rate for each point from \cite{jin12}, and run a corresponding model, assuming constant magnetic parameters 
$\alpha_{\rm B} = 0.18$, $\eta = 0.21$ and $\nu = 1$.
The global disk parameters were chosen to fall roughly in the middle of the range consistent with the observations.
We then proceed and tune the parameter $\alpha_{\rm B}$ for each model, so that at the optical depth $\tau_{\rm cor}$  the average temperature yields $T_{\rm avg}$, both values taken from the observational data.
For some models, it is not possible, but for most (44 objects from the sample of 51) such value of $\alpha_{\rm B}$ does exist.
Both fixed-parameter and fine-tuned models are shown in Fig.~\ref{fig:jin_profiles}, 
where different colors display different disk global parameters. 
The observed  cases of $T_{\rm avg}$ versus  $\tau_{\rm cor}$  are marked by filled circles, while the same parameters as an outcome of our adjusted models are given by crosses. Additionally, by solid thin lines, we plot the averaged temperature integrated from the surface down to the local optical depth given on the x-axis of our graph. Thick solid lines display the extensions of thermally unstable zones. High accretion rate cases at the top do not indicate TI. It is clear from the figure that tuning the magnetic parameters does not result in much change, which means that the fixed values are a good approximation for this set of observations.

In most cases, the observed optical depth of the warm corona is shallower than the
$\tau_{\rm cor}$ derived from our model and indicated by crosses in the figure, with exception to a few models with the lowest accretion rate, where it seems to be deeper than predicted. For a range of models, the location of the warm corona is very coincident with the bottom rather than the top of the TI, and the locations of the two exhibit some correlation in that range. Above some threshold, our model does not predict the occurrence of the TI yet the observed optical depth falls short compared to the $\tau_{\rm cor}$ from our model.


\section{Discussion}
\label{sec:dis}

We computed sample of models of the disk/corona vertical structure, where both layers are physically bounded by magnetic field heating and Compton and free-free cooling. 
Our approach to MRI description is parameterized to adjust the results obtained from 
simulations as explained by GR20. The warm optically thick corona is formed self consistently on the top of an accretion disk
for the range of global disk and magnetic parameters. 
In general the optical depth of warm corona is larger for higher accretion rate and 
higher magnetization. Nevertheless, the data show that even models of moderate accretion rate reproduce observations when magnetic parameters are adjusted. 

Magnetic heating does not remove { the classical TI under constant gas pressure}
presented in regions of 
high temperatures, of the order of 1 keV, and high densities of the order of 10$^{12}$ cm$^{-3}$, cooled by radiation processes. Interestingly, such TI is created only for Compton and free-free absorption/emission. 
{ When magnetic heating with simple description of MRI and recconection is taken into account as a main source of gas heating the disk becomes dominated by magnetic pressure which acts as a freezer
for TI zone. In order to check this behavior, stability parameter should be computed 
under constant gas plus magnetic pressure. 
In such a case, classical TI exists with negative slope of temperature versus ionization parameter, but stability parameter under constant gas plus magnetic pressure is always positive. To prove 
if such gas is stable, the timescales for TI frozen into magnetic field should be estimated and we plan to do it in the forthcoming paper. In the all previous approaches to the ionized disk atmospheres, the ionization structure was always solved only up to the depth of photosphere, i.e. to the lower branch which was stabilized due to atomic cooling and heating.}
Many papers showing such stability curve never run the computations deeper in the disk structure where the energy is dissipated. 
In our approach, the magnetic heating which is generated at each point of a vertical structure of the disk and corona,
{
provides to the normal behavior of the gas i.e the cold disk temperature increases when going towards the equatorial plane, 
which is explicitly presented in panel \texttt{a} of Fig.~\ref{fig:cute}.
}

Compton and free-free processes fully account for the extension of stability curve 
from the hot layers on the corona surface, down to the corona base at $\tau_{\rm cor}$. 
 Ionization/recombination may increase the unstable zone, but this should be 
checked in the future studies. TI may impact the gas evolution in thermal timescales. 
Future, time-dependent simulations may show the impact of TI on the warm corona gas.


{ For all analyzed models the structure was stable when the isobaric constraint
of constant gas plus magnetic pressure was assumed. Of course, such treatment of a magnetic field only holds true in idealized case, assuming that field is frozen into matter, and there are no significant pressure gradients along the flux tube. 
If there is a pressure change, the magnetic structure would need to inflate as a whole, without any escape of the matter.
Practically, magnetic structures in the corona are tangled and can be considered as pressurized but open.
If a thermal runaway or collapse occurred in a flux tube, matter could be pushed or sucked into the flux tube in order to maintain the pressure equilibrium.
In such case, even in presence of a strong field, the original criterion of constant gas pressure would hold and strong TI would occur within separate flux tubes, forming prominence-like structures.
If we searched for conditions to obtain an uniform, warm corona (and not clumpy prominence's), the density $n_0$ (Eq.~\ref{eq:n0}) would still be a sensible order-of-magnitude limitation.}


\subsection{Observations predictions and modeling implications}

As we demonstrated in our paper, observations follow our model of warm corona, 
heated by MRI. All of the data points are in the range $T=$0.01-1\,keV and $\tau_{\rm es}=$2-50. Corona temperature appears to be strongly correlated with coronal magnetic field (particularly, its spatial gradient).
Accretion rate, on the other hand, seems to increase the optical depth of the corona, at the slight cost of average temperature.

Our numerical model has several weaknesses. 
The largest limitation to our model is that it is a static and one-dimensional. That implies that the corona assumed must be layer-like and clumpy, and unstable, but optically thick clumps cannot be studied. 
With the degree that the models are affected by the thermal instability, according to our study, we cannot definitely exclude at this point that the clumpy corona can also reproduce the observations.
It is possible that a dynamical model could resolve the issue of a local instability as long as the overall optically thick disk structure survives.

It is also possible, that thin-disk approximation is not well-suited to the problem, as there is growing evidence that multi-layer accretion, where the accretion of disk and corona are decoupled, may be more realistic \citep{2019ApJ...884L..37L,2022MNRAS.tmp.1281W}.
We are unable to treat such an accretion mode with our approach.

Another limitation is that we do not obtain the spectrum from our model.
While quantities such as optical depth and temperature can be fitted to spectra and we compare the values with our model, there is some degeneracy, and the best way to verify the correctness of our predictions would be to at least compare the spectral parameter $\Gamma$ which is the slope of the power law usually fitted to the data.

Despite those limitations, our model allows to determine the location and properties of the TI. That being said, we cannot exclude that there is a coincidence of occurrence 
of TI i our models and the warm corona in nature.
One peculiar possibility is that the warm corona could actually be driven by the TI.
Instead of one layer, many scatterings through more and less dense areas could be more effective than a single, uniform layer in producing Comptonization spectrum.
More investigation is needed using dynamic models that allow the unstable, clumpy medium to develop (including the magnetic field as essential part to maintain the physical structure of the clumps) and followed by radiative transfer calculations, to prove whether such structure can indeed create warm Comptonization signatures.

If we assume the opposite but (as far as we consider) more likely alternative that the clumpy medium does not contribute to the observed spectrum, another interesting possibility emerges.
It could be that the corona exists at the verge of the instability. The new warm matter is constantly inflowing, and once it reaches the unstable conditions, it collapses into clumps that are ejected or (more likely) fall back into the disk core.
If the condensation timescale is long compared to the dynamical timescale, the matter could exist at the brink of the instability long enough to continuously Comptonizes the soft photons. 
This could act as a stabilizing mechanism and explain why the observed properties of the warm corona seem to be clustered in a narrow range of the parameter space.
While we cannot distinguish which of these scenarios is closer to the truth, the striking coincidence of the warm corona with TI is hard to miss out.

%
\section{Conclusions}
\label{sec:con}

We clearly demonstrated that magnetic heating can produce warm optically thick corona 
above accretion disk in AGN. Obtained theoretical values of coronal temperatures and optical depths agree with those observed for the best known sources. Magnetic viscosity 
parameter increases the coronal temperature and optical thickens. 
Magnetic heating does not remove the { local thermal instability of the corona (TI) caused by radiation processes 
in the warm ($\sim$1.0 keV), and dense (above ~10$^{12}$ cm$^{-3}$) gas}. Classical TI operates for accretion rates lower than 0.1 in Eddington units, and does not depend much on 
magnetic viscosity, nevertheless always the combination of those parameters fully determines instability strip. { When magnetic heating is taken into account, 
stability parameter computed under constant gas plus magnetic pressure is always positive, suggesting that thermally unstable zone in the classical way, becomes a kind of frozen into magnetic field and can produce interesting observational features 
including soft X-ray excess.} 

The case of AGN differs from GBHB, where TI occurs for 
$\dot m < 0,003$ for low $\alpha_{\rm B}$ and $\dot m < 0,05$ for high $\alpha_{\rm B}$
(GR20). Also optical thickness of the warm corona in AGN is relatively higher than in case of GBHB by the factor of 5. 
Interestingly, TI operates only when Compton heating and free-free emission are introduced. It does not require ionization/recombination processes to be included. Nevertheless, we claim here that ionization may influence the extension of the TI zone and we plan to check it in our future work. 

Observations follow TI strip of the warm corona in magnetically supported disk. 
Thermally unstable gas may undergo evolution under thermal timescale, which in principle can be computed in our models. Such theoretical timescale should be compared to the variability measurements of soft corona. We plan to do it in the next step of our 
project. We conclude here, that the TI caused in the gas due to radiative cooling may be a common mechanism affecting the existence of warm corona above accretion disks around black holes across different masses, since we obtain TI in the disk vertical structure in case of AGN, considered here, as well as in case of GBHB presented in our previous paper (GR20).


\begin{acknowledgements}
This research was supported by  Polish National Science Center grants No.
2019/33/N/ST9/02804 and 2021/41/B/ST9/04110.  We acknowledge financial support from the International Space Science Institute (ISSI) through the International Team proposal "Warm coronae in AGN: Observational evidence and physical understanding"\\
\emph{Software}: FORTRAN, Python \citep{python3}, Sympy  \citep{2017-Meurer-sympy},  LAPACK \citep{1990-Anderson-LAPACK} and GNU Parallel \citep{2011-Tange-GNUparallel}.
\end{acknowledgements}


\bibliographystyle{aa}
\bibliography{refs,general,disks,tools}

\begin{thebibliography}{56}
\expandafter\ifx\csname natexlab\endcsname\relax\def\natexlab#1{#1}\fi

\bibitem[{{Adhikari} {et~al.}(2015){Adhikari}, {R{\'o}{\.z}a{\'n}ska},
  {Sobolewska}, \& {Czerny}}]{2015-Adhikari}
{Adhikari}, T.~P., {R{\'o}{\.z}a{\'n}ska}, A., {Sobolewska}, M., \& {Czerny},
  B. 2015, \apj, 815, 83

\bibitem[{Anderson {et~al.}(1990)Anderson, Bai, Dongarra, Greenbaum, McKenney,
  Du~Croz, Hammarling, Demmel, Bischof, \& Sorensen}]{1990-Anderson-LAPACK}
Anderson, E., Bai, Z., Dongarra, J., {et~al.} 1990, in Proceedings of the 1990
  ACM/IEEE Conference on Supercomputing, Supercomputing '90 (Los Alamitos, CA,
  USA: IEEE Computer Society Press), 2--11

\bibitem[{{Ballantyne}(2020)}]{ballantyne2020}
{Ballantyne}, D.~R. 2020, \mnras, 491, 3553

\bibitem[{{Begelman}(2006)}]{2006-Begelman-Bubbles}
{Begelman}, M.~C. 2006, \apj, 643, 1065

\bibitem[{{Begelman} {et~al.}(2015){Begelman}, {Armitage}, \&
  {Reynolds}}]{2015-Begelman}
{Begelman}, M.~C., {Armitage}, P.~J., \& {Reynolds}, C.~S. 2015, \apj, 809, 118

\bibitem[{{Bianchi} {et~al.}(2009){Bianchi}, {Guainazzi}, {Matt}, {Fonseca
  Bonilla}, \& {Ponti}}]{bianchi2009}
{Bianchi}, S., {Guainazzi}, M., {Matt}, G., {Fonseca Bonilla}, N., \& {Ponti},
  G. 2009, \aap, 495, 421

\bibitem[{{Crummy} {et~al.}(2006){Crummy}, {Fabian}, {Gallo}, \&
  {Ross}}]{crummy2006}
{Crummy}, J., {Fabian}, A.~C., {Gallo}, L., \& {Ross}, R.~R. 2006, \mnras, 365,
  1067

\bibitem[{{Done} {et~al.}(2012){Done}, {Davis}, {Jin}, {Blaes}, \&
  {Ward}}]{2012-Done-SXE}
{Done}, C., {Davis}, S.~W., {Jin}, C., {Blaes}, O., \& {Ward}, M. 2012, \mnras,
  420, 1848

\bibitem[{{Field}(1965)}]{field1965}
{Field}, G.~B. 1965, \apj, 142, 531

\bibitem[{{Garc{\'\i}a} {et~al.}(2019){Garc{\'\i}a}, {Kara}, {Walton},
  {Beuchert}, {Dauser}, {Gatuzz}, {Balokovic}, {Steiner}, {Tombesi}, {Connors},
  {Kallman}, {Harrison}, {Fabian}, {Wilms}, {Stern}, {Lanz}, {Ricci}, \&
  {Ballantyne}}]{2019ApJ...871...88G}
{Garc{\'\i}a}, J.~A., {Kara}, E., {Walton}, D., {et~al.} 2019, \apj, 871, 88

\bibitem[{{Gierli{\'n}ski} \& {Done}(2004{\natexlab{a}})}]{gierlinski2004}
{Gierli{\'n}ski}, M. \& {Done}, C. 2004{\natexlab{a}}, \mnras, 349, L7

\bibitem[{{Gierli{\'n}ski} \& {Done}(2004{\natexlab{b}})}]{2004-GierlinskiDone}
{Gierli{\'n}ski}, M. \& {Done}, C. 2004{\natexlab{b}}, \mnras, 349, L7

\bibitem[{{Gronkiewicz} \& {R{\'o}{\.z}a{\'n}ska}(2020)}]{2020-Gronkiewicz}
{Gronkiewicz}, D. \& {R{\'o}{\.z}a{\'n}ska}, A. 2020, \aap, 633, A35

\bibitem[{{Haardt} \& {Maraschi}(1991)}]{1991-HaardtMaraschi}
{Haardt}, F. \& {Maraschi}, L. 1991, \apjl, 380, L51

\bibitem[{{Henyey} {et~al.}(1964){Henyey}, {Forbes}, \& {Gould}}]{Henyey1964}
{Henyey}, L.~G., {Forbes}, J.~E., \& {Gould}, N.~L. 1964, \apj, 139, 306

\bibitem[{{Janiuk} \& {Czerny}(2011)}]{2011-Janiuk}
{Janiuk}, A. \& {Czerny}, B. 2011, \mnras, 414, 2186

\bibitem[{{Jiang} {et~al.}(2014){Jiang}, {Stone}, \& {Davis}}]{2014-Jiang}
{Jiang}, Y.-F., {Stone}, J.~M., \& {Davis}, S.~W. 2014, \apj, 784, 169

\bibitem[{{Jin} {et~al.}(2012){Jin}, {Ward}, {Done}, \& {Gelbord}}]{jin12}
{Jin}, C., {Ward}, M., {Done}, C., \& {Gelbord}, J. 2012, \mnras, 420, 1825

\bibitem[{{Kato} {et~al.}(2008){Kato}, {Fukue}, \& {Mineshige}}]{2008-Kato}
{Kato}, S., {Fukue}, J., \& {Mineshige}, S. 2008, {Black-Hole Accretion Disks
  --- Towards a New Paradigm ---}

\bibitem[{{Keek} \& {Ballantyne}(2016)}]{2016-Keek}
{Keek}, L. \& {Ballantyne}, D.~R. 2016, \mnras, 456, 2722

\bibitem[{{Krolik} {et~al.}(1981){Krolik}, {McKee}, \& {Tarter}}]{krolik81}
{Krolik}, J.~H., {McKee}, C.~F., \& {Tarter}, C.~B. 1981, \apj, 249, 422

\bibitem[{{Lan{\v{c}}ov{\'a}} {et~al.}(2019){Lan{\v{c}}ov{\'a}}, {Abarca},
  {Klu{\'z}niak}, {Wielgus}, {Saḑowski}, {Narayan}, {Schee}, {T{\"o}r{\"o}k},
  \& {Abramowicz}}]{2019ApJ...884L..37L}
{Lan{\v{c}}ov{\'a}}, D., {Abarca}, D., {Klu{\'z}niak}, W., {et~al.} 2019,
  \apjl, 884, L37

\bibitem[{{Laor} {et~al.}(1994){Laor}, {Fiore}, {Elvis}, {Wilkes}, \&
  {McDowell}}]{1994-Laor-1}
{Laor}, A., {Fiore}, F., {Elvis}, M., {Wilkes}, B.~J., \& {McDowell}, J.~C.
  1994, \apj, 435, 611

\bibitem[{{Laor} {et~al.}(1997){Laor}, {Fiore}, {Elvis}, {Wilkes}, \&
  {McDowell}}]{1997-Laor-2}
{Laor}, A., {Fiore}, F., {Elvis}, M., {Wilkes}, B.~J., \& {McDowell}, J.~C.
  1997, \apj, 477, 93

\bibitem[{{Lightman} \& {Eardley}(1974)}]{1974-Lightman}
{Lightman}, A.~P. \& {Eardley}, D.~M. 1974, \apjl, 187, L1

\bibitem[{{Madau}(1988)}]{1988-Madau}
{Madau}, P. 1988, \apj, 327, 116

\bibitem[{{Magdziarz} {et~al.}(1998){Magdziarz}, {Blaes}, {Zdziarski},
  {Johnson}, \& {Smith}}]{1998-Magdziarz}
{Magdziarz}, P., {Blaes}, O.~M., {Zdziarski}, A.~A., {Johnson}, W.~N., \&
  {Smith}, D.~A. 1998, \mnras, 301, 179

\bibitem[{{Matt} {et~al.}(2014){Matt}, {Marinucci}, {Guainazzi}, {Brenneman},
  {Elvis}, {Lohfink}, {Ar{\`e}valo}, {Boggs}, {Cappi}, {Christensen}, {Craig},
  {Fabian}, {Fuerst}, {Hailey}, {Harrison}, {Parker}, {Reynolds}, {Stern},
  {Walton}, \& {Zhang}}]{2014-Matt}
{Matt}, G., {Marinucci}, A., {Guainazzi}, M., {et~al.} 2014, \mnras, 439, 3016

\bibitem[{{Mehdipour} {et~al.}(2011){Mehdipour}, {Branduardi-Raymont},
  {Kaastra}, {Petrucci}, {Kriss}, {Ponti}, {Blustin}, {Paltani}, {Cappi},
  {Detmers}, \& {Steenbrugge}}]{2011-Mehdipour-Mrk509}
{Mehdipour}, M., {Branduardi-Raymont}, G., {Kaastra}, J.~S., {et~al.} 2011,
  \aap, 534, A39

\bibitem[{{Mehdipour} {et~al.}(2015){Mehdipour}, {Kaastra}, {Kriss}, {Cappi},
  {Petrucci}, {Steenbrugge}, {Arav}, {Behar}, {Bianchi}, {Boissay}, {Brand
  uardi-Raymont}, {Costantini}, {Ebrero}, {Di Gesu}, {Harrison}, {Kaspi}, {De
  Marco}, {Matt}, {Paltani}, {Peterson}, {Ponti}, {Pozo Nu{\~n}ez}, {De Rosa},
  {Ursini}, {de Vries}, {Walton}, \& {Whewell}}]{2015-Mehdipour}
{Mehdipour}, M., {Kaastra}, J.~S., {Kriss}, G.~A., {et~al.} 2015, \aap, 575,
  A22

\bibitem[{Meurer {et~al.}(2017)Meurer, Smith, Paprocki, \v{C}ert\'{i}k,
  Kirpichev, Rocklin, Kumar, Ivanov, Moore, Singh, Rathnayake, Vig, Granger,
  Muller, Bonazzi, Gupta, Vats, Johansson, Pedregosa, Curry, Terrel,
  Rou\v{c}ka, Saboo, Fernando, Kulal, Cimrman, \& Scopatz}]{2017-Meurer-sympy}
Meurer, A., Smith, C.~P., Paprocki, M., {et~al.} 2017, PeerJ Computer Science,
  3, e103

\bibitem[{{Middei} {et~al.}(2018){Middei}, {Bianchi}, {Cappi}, {Petrucci},
  {Ursini}, {Arav}, {Behar}, {Brand uardi-Raymont}, {Costantini}, {De Marco},
  {Di Gesu}, {Ebrero}, {Kaastra}, {Kaspi}, {Kriss}, {Mao}, {Mehdipour},
  {Paltani}, {Peretz}, \& {Ponti}}]{2018-Middei}
{Middei}, R., {Bianchi}, S., {Cappi}, M., {et~al.} 2018, \aap, 615, A163

\bibitem[{{Middei} {et~al.}(2019){Middei}, {Bianchi}, {Petrucci}, {Ursini},
  {Cappi}, {De Marco}, {De Rosa}, {Malzac}, {Marinucci}, {Matt}, {Ponti}, \&
  {Tortosa}}]{2019-Middei}
{Middei}, R., {Bianchi}, S., {Petrucci}, P.~O., {et~al.} 2019, \mnras, 483,
  4695

\bibitem[{{Page} {et~al.}(2004){Page}, {Schartel}, {Turner}, \&
  {O'Brien}}]{page2004}
{Page}, K.~L., {Schartel}, N., {Turner}, M.~J.~L., \& {O'Brien}, P.~T. 2004,
  \mnras, 352, 523

\bibitem[{{Pessah} \& {Psaltis}(2005)}]{2005-PessahPsaltis}
{Pessah}, M.~E. \& {Psaltis}, D. 2005, \apj, 628, 879

\bibitem[{{Petrucci} {et~al.}(2020){Petrucci}, {Gronkiewicz}, {Rozanska},
  {Belmont}, {Bianchi}, {Czerny}, {Matt}, {Malzac}, {Middei}, {De Rosa},
  {Ursini}, \& {Cappi}}]{petrucci2020}
{Petrucci}, P.~O., {Gronkiewicz}, D., {Rozanska}, A., {et~al.} 2020, \aap, 634,
  A85

\bibitem[{{Petrucci} {et~al.}(2013){Petrucci}, {Paltani}, {Malzac}, {Kaastra},
  {Cappi}, {Ponti}, {De Marco}, {Kriss}, {Steenbrugge}, {Bianchi},
  {Branduardi-Raymont}, {Mehdipour}, {Costantini}, {Dadina}, \&
  {Lubi{\'n}ski}}]{2013-Petrucci-Mrk509}
{Petrucci}, P.-O., {Paltani}, S., {Malzac}, J., {et~al.} 2013, \aap, 549, A73

\bibitem[{{Petrucci} {et~al.}(2018){Petrucci}, {Ursini}, {De Rosa}, {Bianchi},
  {Cappi}, {Matt}, {Dadina}, \& {Malzac}}]{2018-Petrucci}
{Petrucci}, P.~O., {Ursini}, F., {De Rosa}, A., {et~al.} 2018, \aap, 611, A59

\bibitem[{{Piconcelli} {et~al.}(2005){Piconcelli}, {Jimenez-Bail{\'o}n},
  {Guainazzi}, {Schartel}, {Rodr{\'{\i}}guez-Pascual}, \&
  {Santos-Lle{\'o}}}]{2005-Piconcelli-SXE}
{Piconcelli}, E., {Jimenez-Bail{\'o}n}, E., {Guainazzi}, M., {et~al.} 2005,
  \aap, 432, 15

\bibitem[{{Porquet} {et~al.}(2018){Porquet}, {Reeves}, {Matt}, {Marinucci},
  {Nardini}, {Braito}, {Lobban}, {Ballantyne}, {Boggs}, {Christensen},
  {Dauser}, {Farrah}, {Garcia}, {Hailey}, {Harrison}, {Stern}, {Tortosa},
  {Ursini}, \& {Zhang}}]{2018-Porquet}
{Porquet}, D., {Reeves}, J.~N., {Matt}, G., {et~al.} 2018, \aap, 609, A42

\bibitem[{{Pounds} {et~al.}(1987){Pounds}, {Stanger}, {Turner}, {King}, \&
  {Czerny}}]{1987-Pounds-Mkn335}
{Pounds}, K.~A., {Stanger}, V.~J., {Turner}, T.~J., {King}, A.~R., \& {Czerny},
  B. 1987, \mnras, 224, 443

\bibitem[{{R{\'o}{\.z}a{\'n}ska}(1999)}]{1999-Rozanska-Conduction}
{R{\'o}{\.z}a{\'n}ska}, A. 1999, \mnras, 308, 751

\bibitem[{{R\'o\.za\'nska} \& {Czerny}(1996)}]{1996-RozanskaCzerny}
{R\'o\.za\'nska}, A. \& {Czerny}, B. 1996, \actaa, 46, 233

\bibitem[{{R{\'o}{\.z}a{\'n}ska} {et~al.}(1999){R{\'o}{\.z}a{\'n}ska},
  {Czerny}, {{\.Z}ycki}, \& {Pojma{\'n}ski}}]{1999-Rozanska}
{R{\'o}{\.z}a{\'n}ska}, A., {Czerny}, B., {{\.Z}ycki}, P.~T., \&
  {Pojma{\'n}ski}, G. 1999, \mnras, 305, 481

\bibitem[{{R{\'o}{\.z}a{\'n}ska} {et~al.}(2015){R{\'o}{\.z}a{\'n}ska},
  {Malzac}, {Belmont}, {Czerny}, \& {Petrucci}}]{rozanska15}
{R{\'o}{\.z}a{\'n}ska}, A., {Malzac}, J., {Belmont}, R., {Czerny}, B., \&
  {Petrucci}, P.-O. 2015, \aap, 580, A77

\bibitem[{{Salvesen} {et~al.}(2016){Salvesen}, {Simon}, {Armitage}, \&
  {Begelman}}]{2016-Salvesen}
{Salvesen}, G., {Simon}, J.~B., {Armitage}, P.~J., \& {Begelman}, M.~C. 2016,
  \mnras, 457, 857

\bibitem[{{Shakura} \& {Sunyaev}(1973)}]{1973-ShakuraSunyaev}
{Shakura}, N.~I. \& {Sunyaev}, R.~A. 1973, \aap, 24, 337

\bibitem[{{Shakura} \& {Sunyaev}(1976)}]{1976-Shakura}
{Shakura}, N.~I. \& {Sunyaev}, R.~A. 1976, \mnras, 175, 613

\bibitem[{Shibazaki \& Hōshi(1975)}]{1975-Shibazaki}
Shibazaki, N. \& Hōshi, R. 1975, Progress of Theoretical Physics, 54, 706

\bibitem[{{Szuszkiewicz} \& {Miller}(1997)}]{szuszkiewicz}
{Szuszkiewicz}, E. \& {Miller}, J.~C. 1997, \mnras, 287, 165

\bibitem[{Tange(2011)}]{2011-Tange-GNUparallel}
Tange, O. 2011, ;login: The USENIX Magazine, 36, 42

\bibitem[{{Ursini} {et~al.}(2018){Ursini}, {Petrucci}, {Matt}, {Bianchi},
  {Cappi}, {Dadina}, {Grandi}, {Torresi}, {Ballantyne}, {De Marco}, {De Rosa},
  {Giroletti}, {Malzac}, {Marinucci}, {Middei}, {Ponti}, \&
  {Tortosa}}]{2018-Ursini}
{Ursini}, F., {Petrucci}, P.~O., {Matt}, G., {et~al.} 2018, \mnras, 478, 2663

\bibitem[{Van~Rossum \& Drake(2009)}]{python3}
Van~Rossum, G. \& Drake, F.~L. 2009, Python 3 Reference Manual (Scotts Valley,
  CA: CreateSpace)

\bibitem[{{Walter} \& {Fink}(1993)}]{walter93}
{Walter}, R. \& {Fink}, H.~H. 1993, \aap, 274, 105

\bibitem[{{Walton} {et~al.}(2013){Walton}, {Nardini}, {Fabian}, {Gallo}, \&
  {Reis}}]{walton2013}
{Walton}, D.~J., {Nardini}, E., {Fabian}, A.~C., {Gallo}, L.~C., \& {Reis},
  R.~C. 2013, \mnras, 428, 2901

\bibitem[{{Wielgus} {et~al.}(2022){Wielgus}, {Lan{\v{c}}ov{\'a}}, {Straub},
  {Klu{\'z}niak}, {Narayan}, {Abarca}, {R{\'o}{\.z}a{\'n}ska}, {Vincent},
  {T{\"o}r{\"o}k}, \& {Abramowicz}}]{2022MNRAS.tmp.1281W}
{Wielgus}, M., {Lan{\v{c}}ov{\'a}}, D., {Straub}, O., {et~al.} 2022, \mnras
  [\eprint[arXiv]{2202.08831}]

\end{thebibliography}


\end{document}